# New methods of testing Lorentz violation in electrodynamics

Michael E. Tobar[1], Peter Wolf[2, 3], Alison Fowler[1], John G. Hartnett[1]

[1]*University of Western Australia, School of Physics, Crawley 6009 WA, Australia*
[2]*BNM-SYRTE, Observatoire de Paris, 61 Av. de l'Observatoire, 75014 Paris, France*
[3]*Bureau International des Poids et Mesures, Pavillon de Breteuil, 92312 Sèvres CEDEX, France*

We investigate experiments that are sensitive to the scalar and parity-odd coefficients for Lorentz violation in the photon sector of the Standard Model Extension (SME). We show that of the classic tests of special relativity, Ives-Stilwell (IS) experiments are sensitive to the scalar coefficient, but at only parts in $10^5$ for the state-of-the-art experiment. We then propose asymmetric Mach-Zehnder interferometers with different electromagnetic properties in the two arms, including recycling techniques based on travelling wave resonators to improve the sensitivity. With present technology we estimate that the scalar and parity odd coefficients may be measured at sensitivity better than parts in $10^{11}$ and $10^{15}$ respectively.

## I. INTRODUCTION

The postulate of Lorentz Invariance (LI) is at the heart of special and general relativity and therefore one of the cornerstones of modern physics. The central importance of this postulate has motivated tremendous work to experimentally test LI with ever increasing precision[1]. Additionally, many unification theories (e.g. string theory or loop gravity) are expected to violate LI at some level,[2-4] which further motivates experimental searches for such violations.

Numerous test theories that allow the modelling and interpretation of experiments that test LI have been developed. Kinematical frameworks[5,6] postulate a simple parameterisation of the Lorentz transformations with experiments setting limits on the deviation of those parameters from their values in special relativity. A more fundamental approach is offered by theories that parameterize the coupling between gravitational and non-gravitational fields (e.g. THεμ formalisms[1,7]). Formalisms based on string theory [2, 3] have the advantage of being well motivated by theories of physics that are at present good candidates for a unification of gravity and the other fundamental forces of nature. Fairly recently a general Lorentz violating extension of the standard model of particle physics (Standard Model Extension, SME) has been developed[8-10] whose Lagrangian includes all parameterised Lorentz violating terms that can be formed from known fields. Many of the theories mentioned above are included as special cases of the SME[11]. In this paper we restrict our attention to the photon sector of the SME. Within this framework we analyse past experiments that can be shown to set limits on SME parameters that have not been determined previously, and propose new experiments that could significantly improve those limits.

As shown in[11] the photon sector of the SME can be expressed in the form of modified source free Maxwell equations, which take their familiar form

$$\nabla . \boldsymbol{D} = 0 \tag{1a}$$
$$\nabla . \boldsymbol{B} = 0 \tag{1b}$$
$$\nabla \times \boldsymbol{E} + \partial \boldsymbol{B} / \partial t = 0 \tag{1c}$$
$$\nabla \times \boldsymbol{H} - \partial \boldsymbol{D} / \partial t = 0 \tag{1d}$$

but with modified definitions of $\boldsymbol{D}$ and $\boldsymbol{H}$

$$\begin{bmatrix} \mathbf{D} \\ \mathbf{H} \end{bmatrix} = \begin{bmatrix} \varepsilon_0(\vec{\varepsilon}_r + \kappa_{DE}) & \sqrt{\frac{\varepsilon_0}{\mu_0}}\kappa_{DB} \\ \sqrt{\frac{\varepsilon_0}{\mu_0}}\kappa_{HE} & \mu_0^{-1}(\vec{\mu}_r^{-1} + \kappa_{HB}) \end{bmatrix} \begin{bmatrix} \mathbf{E} \\ \mathbf{B} \end{bmatrix} \tag{2}$$

Here $\kappa_{DE}$, $\kappa_{DB}$, $\kappa_{HE}$ and $\kappa_{HB}$ are all 3 x 3 matrices, which parameterize possible Lorentz violating terms as described in[11]. If we suppose the medium of interest has general magnetic or dielectric properties, then $\vec{\varepsilon}_r$ and $\vec{\mu}_r$ are also 3 x 3 matrices. In vacuum $\vec{\varepsilon}_r$ and $\vec{\mu}_r$ are identity matrices. For experimental tests it is convenient to further define linear combinations of the $\kappa$ coefficients

$$\begin{aligned} (\tilde{\kappa}_{e+})^{jk} &= \tfrac{1}{2}(\kappa_{DE} + \kappa_{HB})^{jk}, \\ (\tilde{\kappa}_{e-})^{jk} &= \tfrac{1}{2}(\kappa_{DE} - \kappa_{HB})^{jk} - \tfrac{1}{3}\delta^{kj}(\kappa_{DE})^{ll}, \\ (\tilde{\kappa}_{o+})^{jk} &= \tfrac{1}{2}(\kappa_{DB} + \kappa_{HE})^{jk}, \\ (\tilde{\kappa}_{o-})^{jk} &= \tfrac{1}{2}(\kappa_{DB} - \kappa_{HE})^{jk}, \\ \tilde{\kappa}_{tr} &= \tfrac{1}{3}(\kappa_{DE})^{ll}. \end{aligned} \tag{3}$$

The first four of these equations define traceless 3 x 3 matrices, while the last defines a single coefficient. All $\tilde{\kappa}$ matrices are symmetric except $\tilde{\kappa}_{o+}$ which is antisymmetric (odd parity). There are 19 independent coefficients of the $\kappa$ tensors, which are generally used to quote and compare experimental results[11-15].

The $\kappa$ tensors in (2) and (3) are frame dependent and consequently vary as a function of the co-ordinate system chosen to analyse a given experiment. In principle they may be constant and non-zero in any frame (e.g. the lab frame). However, any non-zero values are expected to arise from Planck-scale effects in the early Universe. Therefore the components of $\kappa$ should be constant in a cosmological frame (e.g. the one defined by the CMB radiation) or any frame that moves with a constant velocity and shows no rotation with respect to the cosmological one. Consequently the conventionally chosen frame to analyze and compare experiments in the SME is a sun-centred, non-rotating frame as defined in[11]. The general procedure is to express the experimental observable in terms of the $\kappa$ tensors in a suitably chosen experimental frame (e.g. the lab frame) and then to transform the $\kappa$ tensors to the conventional sun-centred frame. This transformation will introduce a time variation of the observable related to the movement of the experiment with respect to the sun-centred frame (typically introducing time variations of sidereal and semi-sidereal periods for an Earth fixed experiment).

So far two types of experiments have been used to set limits on 17 of the 19 independent components of $\tilde{\kappa}$. Polarization measurements of light from distant astrophysical sources have been used to constrain all 10 independent components of $\tilde{\kappa}_{e+}$ and $\tilde{\kappa}_{o-}$ to less than $2 \times 10^{-32}$.[11,12] Modern versions of classical tests of special relativity (Michelson-Morley and Kennedy-Thorndike experiments) using optical[16] or microwave[13, 15, 17] cavities have recently constrained 4 components of $\tilde{\kappa}_{e-}$ to a few parts in $10^{-15}$ and all three independent components of $\tilde{\kappa}_{o+}$ to a few parts in $10^{-11}$.[15] For the time being, this leaves two parameters undetermined, $\tilde{\kappa}_{e-}^{ZZ}$ and $\tilde{\kappa}_{tr}$.

So far all cavity experiments analysed in the SME[13,15,16] have been fixed in the laboratory, and therefore their orientation in the sun-centred frame varied only with the rotation of the Earth. This induces symmetry with respect to the Earth's rotation axis, which makes these experiments insensitive to $\tilde{\kappa}_{e-}^{ZZ}$. New cavity experiments that rotate in the lab are under way[18] and are expected to measure all 5 independent components of $\tilde{\kappa}_{e-}$ (including $\tilde{\kappa}_{e-}^{ZZ}$) with an uncertainty of $10^{-16}$ or less. This will then leave only $\tilde{\kappa}_{tr}$ undetermined. Experiments that are sensitive to that parameter are inherently difficult as $\tilde{\kappa}_{tr}$ characterises the isotropic part of $\kappa_{DE}$ and $\kappa_{HB}$ in the sun centred frame, so any change in orientation of the experiment does not affect the $\tilde{\kappa}_{tr}$ dependence of the observable. Therefore, modulation of the $\tilde{\kappa}_{tr}$ dependence arises only from first or second order boost (depending on the experiment), i.e. from terms that contribute to $\tilde{\kappa}_{tr}$ via the velocity of the experiment in the sun-centred frame. In this paper we examine existing experiments showing that such a mechanism can indeed lead to sensitivity to $\tilde{\kappa}_{tr}$, and propose new experiments that could determine $\tilde{\kappa}_{tr}$ at a level of better than $10^{-11}$.

Throughout this paper we concentrate only on $\tilde{\kappa}_{tr}$, and where appropriate the components of the odd parity tensor $\tilde{\kappa}_{0+}$. Consequently, we assume that the matter sector of the SME conforms to Lorentz symmetry. Also, we will discuss sensitivities to $\tilde{\kappa}_{tr}$ which are of order $10^{-5}$ – $10^{-11}$, and sensitivities to $\tilde{\kappa}_{0+}$ of order $10^{-15}$. In some instances the sensitivity is several orders of magnitude worse than the best present limits for the other photon parameters (see above), and therefore, we will set all other photon parameters to zero. In other instances we just set the 10 polarization dependent components to zero as they have been measured to parts in $10^{32}$. In all cases these substitutions simplify the resulting expressions without affecting our conclusions.

In section II we analyse existing Lorentz invariance tests and identify a class of experiments that turn out to be sensitive to $\tilde{\kappa}_{tr}$. We explicitly model one of those experiments (at present, the most sensitive one) in the SME and derive an order of magnitude estimate for the limit that experiment sets on $\tilde{\kappa}_{tr}$. In section III we propose new interferometric experiments and estimate their sensitivity to $\tilde{\kappa}_{tr}$ and $\tilde{\kappa}_{0+}$. We show that such experiments should improve on the limit of $\tilde{\kappa}_{tr}$ derived in section II by several orders of magnitude. In section IV we discuss the possibility of further improving the sensitivity using asymmetric high-Q resonators. We summarise our results and conclude in section V.

## II. LIMITS ON $\tilde{\kappa}_{tr}$ FROM PREVIOUS EXPERIMENTS

Classical tests of special relativity (or Lorentz invariance) are usually grouped into three classes: Michelson-Morley (MM)[19], Kennedy-Thorndike (KT)[20] and Ives-Stilwell (IS) experiments.[21] The latter are sometimes referred to as Doppler, clock-comparison or one-way speed of light experiments. As mentioned above, MM and KT experiments have already been used to set stringent limits on a number of SME parameters in the photon sector. In this section we analyse IS experiments in the photon sector of the SME and derive the limit on $\tilde{\kappa}_{tr}$ obtained from the most sensitive such experiments to date[22].

In the original IS experiment[21] hydrogen atoms were moving at $v_{at}/c \approx 0.005$ ($c$ = 299 792 458 m/s) with respect to the laboratory. The Doppler shifted frequencies of the $H_\beta$ line were measured in parallel ($\nu_p$) and anti-parallel ($\nu_a$) to the direction of motion of the atom (see fig. 1). The results are then used to determine whether the combination $\nu_p\nu_a/\nu_0^2 = 1$ as required in special relativity (where $\nu_0$ is the frequency for $v_{at}$=0).

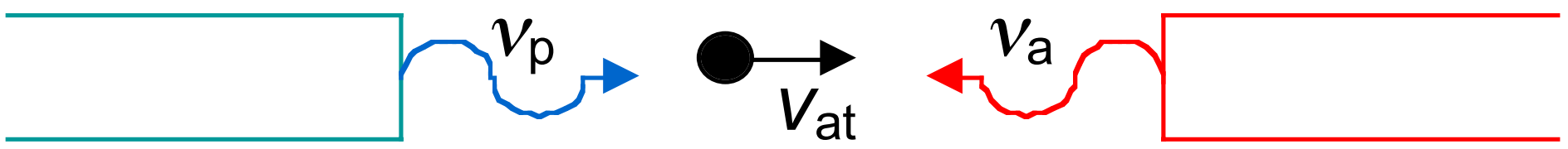

Figure 1: Principle of the Ives-Stilwell experiment. The Doppler shifted resonant frequencies of a moving atom are measured in the parallel $\nu_p$, and anti-parallel $\nu_a$ directions.

Modern experiments of this type include Mössbauer rotor experiments[23], two photon and saturation spectroscopy experiments on atoms or ions,[22,24,25] or experiments that compare atomic clocks over large distances.[26-28] Broadly speaking, all such experiments search for an anisotropy of the first order Doppler shift (or the one-way phase velocity of light) although some caution is required when physically interpreting such statements (in particular one needs to unambiguously define the meaning of "one-way light velocity"). More rigorously, all such experiments have been interpreted in the theoretical framework of Robertson, Mansouri and Sexl (RMS)[5,6] in which they set limits on the parameter $\alpha_{RMS}$ (one of the three fundamental RMS parameters, the other two being determined by MM and KT experiments). Limits on $\alpha_{RMS}$ are obtained from[25,27] that limit $|\alpha_{RMS} + 1/2| \leq 8 \times 10^{-7}$ and, more recently, by[22] which provides the best limit to date $|\alpha_{RMS} + 1/2| \leq 2.2 \times 10^{-7}$.

As shown in[9, 11] (see also below) the modified Maxwell equations (1), (2) of the SME lead to effects on light propagation, in particular, dependence of the phase velocity of the light on polarization and direction of propagation. This suggests that IS experiments are sensitive to the SME. It turns out that they are in fact sensitive to the at-present undetermined parameter $\tilde{\kappa}_{tr}$ via the difference of the phase velocities of the signals travelling in opposite directions in a lab frame moving with respect to the sun centred frame. In contrast, MM and KT experiments always measure or compare return travel times of light signals, hence the phase velocity difference between the opposing directions cancels in the observables of those experiments, which makes them insensitive to $\tilde{\kappa}_{tr}$. In the next subsections we explicitly derive the sensitivity to $\tilde{\kappa}_{tr}$ for the most accurate IS experiment to date[22].

### A. Vacuum light propagation in the SME

Here we consider solutions to equations (1) and (2) in vacuum i.e. with the tensors $\vec{\varepsilon}_r$ and $\vec{\mu}_r$ in (2) being identity matrices. The more general case of magnetic and/or dielectric materials in the SME is treated in section III. Detailed calculations deriving the plane wave solutions in vacuum in the SME can be found in[9, 11]. Here we recall only the principles and results of those calculations that are relevant to our purpose.

We start with the standard *ansatz*

$$\mathbf{E} = \mathbf{E}_0 e^{i(\beta z - \omega t)}$$
$$\mathbf{B} = \mathbf{B}_0 e^{i(\beta z - \omega t)} \tag{4}$$

for a plane wave propagating in the positive $z$ direction with frequency $\omega$ and wave number $\beta$. Substituting (4) into (1c), solving for $\boldsymbol{B}$, then substituting the result into (2) and (1d) results in a set of three linear coupled equations that can be solved for $\boldsymbol{E}$[9,11]. A non-trivial solution is found only under the condition

$$\frac{\beta_\pm}{k_o} = 1 - \rho \mp \sigma \tag{5}$$

where the ± signs refer to two fundamental modes of propagation $\mathbf{E}_\pm$ and $k_o = \omega\sqrt{\mu_o \varepsilon_o}$ is the propagation constant of a plane wave in vacuum in the absence of Lorentz violation, with

$$\begin{aligned} \rho &= -\frac{1}{4}\left(\kappa_{DE}^{xx} - \kappa_{HB}^{xx} + \kappa_{DE}^{yy} - \kappa_{HB}^{yy} + 2\kappa_{DB}^{xy} + 2\kappa_{HE}^{xy}\right) \\ \sigma^2 &= \left(\frac{1}{4}\left(\kappa_{DB}^{xx} - \kappa_{HE}^{xx}\right) - \frac{1}{4}\left(\kappa_{DB}^{yy} - \kappa_{HE}^{yy}\right) - \frac{1}{2}\left(\kappa_{DE}^{xy} + \kappa_{HB}^{xy}\right)\right)^2 \\ &+ \left(-\frac{1}{4}\left(\kappa_{DE}^{xx} + \kappa_{HB}^{xx}\right) + \frac{1}{4}\left(\kappa_{DE}^{yy} + \kappa_{HB}^{yy}\right) - \frac{1}{2}\left(\kappa_{DB}^{xy} - \kappa_{HE}^{xy}\right)\right)^2 \end{aligned} \tag{6}$$

to first order in $\kappa$. The dispersion relation (5) characterizes the propagation of two fundamental modes with the general solution for arbitrary polarization being a superposition of those two modes. By finding the explicit solutions for the two modes[9,11] it can be shown that, to first order in $\kappa$, the fields $\boldsymbol{E}_\pm$ are perpendicular to each other and also perpendicular to the direction of propagation. In the absence of Lorentz violation all $\kappa$ components vanish, $\rho = \sigma = 0$, and (5) reduces to the usual dispersion relation in vacuum.

### B. The Ives-Stilwell experiment in the photon sector of the SME

The most recent version[22] of the IS experiment is in principle very similar to the original experiment (c.f. Fig. 1). It uses collinear saturation spectroscopy on $^7Li^+$ ions moving at $v_{at}/c = 0.064$ in the heavy-ion storage ring TSR in Heidelberg. A closed two level transition at $\nu_0 = 5.46 \times 10^{14}$ Hz is excited by two iodine stabilised lasers, which are tuned to the Doppler shifted transition frequencies ($\nu_p$ and $\nu_a$). The observed fluorescence signal of the ions shows a Lamb dip characteristic of saturation spectroscopy when both lasers are resonant with the respective Doppler shifted frequencies. Combining the two frequencies the experiment then determines

$$\frac{\nu_p \nu_a}{\nu_0^2} = 1 + \varepsilon_{LV}(t) \tag{7}$$

where $\varepsilon_{LV}$ is a term, in general time varying, that characterises possible Lorentz violation in a given theoretical framework. In special relativity $\varepsilon_{LV} = 0$, and the experiment by Saathoff et al.[22] sets a limit of $\varepsilon_{LV} < 1.8 \times 10^{-9}$. They interpret their result in the theoretical framework of RMS[5,6] in which $\varepsilon_{LV} = 2(\alpha_{RMS} + 1/2)(\mathbf{v}^2_{at} + 2\,\mathbf{v}_{lab}\cdot\mathbf{v}_{at})/c^2$ where $\mathbf{v}_{lab}$ is the velocity of the laboratory in the preferred frame (generally taken as the rest frame of the cosmic microwave background). They obtain a limit of $(\alpha_{RMS} + 1/2) < 2.2 \times 10^{-7}$ under the assumption that $v_{at} >> v_{lab}$.

To interpret the experiment in the SME we first work in an "experiment" frame which is at rest in the laboratory with the atoms moving along its positive $z$-axis (see fig. 1). In that frame the two laser frequencies are $\nu_p$ and $\nu_a$ respectively. Using (5) we define the phase velocities of the parallel and anti-parallel laser beams

$$c_p \equiv \frac{\omega_p}{\beta_p} = c(1+\rho_p \pm \sigma_p)$$
$$c_a \equiv \frac{\omega_a}{\beta_a} = c(1+\rho_a \pm \sigma_a) \tag{8}$$

with $\rho_p$ and $\sigma_p$ given by (6) and $\rho_a$ and $\sigma_a$ obtained from (6) by changing the sign of all $\kappa_{DB}$ and $\kappa_{HE}$ terms. The ± sign again refers to the two fundamental modes, and all quantities (in particular the $\kappa$ tensors) are defined in the "experiment" frame.

We then calculate the time interval (in the experiment frame) between two successive maxima of the parallel and anti-parallel beams encountered by an atom. These are equal to the Doppler shifted periods in the experiment frame ($1/\nu_p$' and $1/\nu_a$') of the light encountered by the atom:

$$\nu'_p = \nu_p(1 - v_{at}/c_p)$$
$$\nu'_a = \nu_a(1 + v_{at}/c_a) \tag{9}$$

where $v_{at}$ is the atomic velocity in the experiment frame.

To obtain the frequencies ($\nu_p$'' and $\nu_a$'') as seen by the atom (the frequencies absorbed by the atom) we need to transform to the "atom" frame, co-moving with the atom. Using a standard Lorentz transformation (justified below) and imposing that the light absorbed by the atom be resonant with the chosen transition (i.e. setting $\nu_p$'' = $\nu_a$'' = $\nu_0$) we obtain

$$\nu_p = \nu_0 \frac{\sqrt{1 - v_{at}^2/c^2}}{1 - v_{at}/c_p} \quad \text{and} \quad \nu_a = \nu_0 \frac{\sqrt{1 - v_{at}^2/c^2}}{1 + v_{at}/c_a} \tag{10}$$

which relates the rest frequency $\nu_0$ of the atomic transition to the measured frequencies of the two lasers on resonance ($\nu_p$ and $\nu_a$).

To obtain expression (10) we have used a standard Lorentz transformation to transform the atomic transition frequency from the atom to the experiment frame. That is because we assume throughout this work that the matter sector is Lorentz invariant. Under this assumption the atomic transition frequency is only affected by a violation of the photon sector, via radiative corrections (e.g. Lamb shift for optical transitions in Hydrogen). However, such corrections only amount to parts in $10^6$ of the transition frequency,[43,44] hence a Lorentz violating modification of those corrections would lead to a negligible effect when compared to the leading order Lorentz violating terms in $v_{at}/c_{p/a}$ in (10) ($v_{at}/c \approx 0.06$).

Combining the two equations of (10) and keeping only first order terms in $v_{at}/c$ we finally obtain an expression for the observable in[22]

$$\frac{\nu_a \nu_p}{\nu_0^2} = 1 + \frac{v_{at}}{c_p} - \frac{v_{at}}{c_a} \tag{11}$$

where the Lorentz violating terms are contained in the phase velocities $c_a$ and $c_p$ defined in (8), and expressed using (6) in the experiment frame (with all $\kappa$ tensors defined in the experiment frame).

We use the relationship $\kappa_{DE} = -\kappa_{HE}^{T}$ [11] and transform the $\kappa$ tensors to a laboratory frame as defined in[11] ($x$-axis pointing south, $z$-axis vertically upwards) by a simple rotation, and to the conventional sun-centred frame by a rotation and a boost (equations (30) and (31) from,[11] see also appendix A). As mentioned above, the sensitivities to $\tilde{\kappa}_{tr}$ discussed here are orders of magnitude worse than the best present limits on all other $\kappa$ components. We therefore set those other coefficients to zero in our final expression in the sun-centred frame, which leaves, after some calculation,

$$\frac{\nu_a \nu_p}{\nu_0^2} = 1 + 4\tilde{\kappa}_{tr}\frac{v_{at}}{c}\left[\begin{array}{l} \cos(\phi)\dfrac{v_\oplus}{c}\left(\begin{array}{l}\cos(\Omega_\oplus T)\big(\sin(\eta)\sin(\chi) - \cos(\eta)\cos(\chi)\sin(\omega_\oplus T_\oplus)\big) \\ + \sin(\Omega_\oplus T)\cos(\chi)\cos(\omega_\oplus T_\oplus)\end{array}\right) \\ + \sin(\phi)\left(\dfrac{v_r}{c} - \dfrac{v_\oplus}{c}\big(\cos(\Omega_\oplus T)\cos(\eta)\cos(\omega_\oplus T_\oplus) + \sin(\Omega_\oplus T)\sin(\omega_\oplus T_\oplus)\big)\right)\end{array}\right]. \quad (12)$$

For the derivation of (12) we have assumed that the atoms move horizontally with a velocity $v_{at}$ in the lab frame and at an angle $\phi$ with respect to local south. The other symbols follow the definitions in[11]: $\eta$ is the declination of the Earth's orbital plane ($\eta \approx 23.4$ °), $\chi$ is the colatitude of the laboratory, $\Omega_\oplus, v_\oplus$ are the angular velocity and speed of the Earth's orbital motion, $\omega_\oplus$ is the angular velocity of the Earth's rotation, and $v_r = r_\oplus \omega_\oplus \sin(\chi)$ is the velocity of the lab due to rotation of the Earth. The times $T$ and $T_\oplus$ respectively are the time since a spring equinox and the time since the lab frame $y$-axis pointed towards 90° right ascension.

The experiment by Saathoff et al.[22] sets an upper limit of $1.8 \times 10^{-9}$ on the deviation of $\nu_p \nu_a / \nu_0^2$ from unity. Using (12) that result implies a limit on $\tilde{\kappa}_{tr}$ of the order $10^{-5}$, with the exact value depending on the angle $\phi$ and the times ($T$ and $T_\oplus$) at which the experiment was carried out. This is, to our knowledge, the only quoted limit on $\tilde{\kappa}_{tr}$ to date. However, the cavity experiments that are directly sensitive to the even parity coefficients could possibly be sensitive to second order in velocity to $\tilde{\kappa}_{tr}$.[40] Since the value of the earth's orbital velocity is of order $10^{-4}$ (with respect to $c$), and the best cavity experiments have a sensitivity of order $10^{-15}$ to the even parity coefficients, the possible sensitivity to $\tilde{\kappa}_{tr}$ would be of order $10^{-7}$. Detailed second order perturbation analysis would be necessary to calculate and verify the exact sensitivity to the coefficient, and has not been achieved to date.

## III. INTERFEROMETER TESTS USING MAGNETIC MATERIALS

In the previous section we have shown that the $\tilde{\kappa}_{tr}$ scalar may be measured by IS experiments, and set a constraint on its value of parts in $10^5$. In general, experiments that are sensitive to the $\tilde{\kappa}_{tr}$ scalar, but suppressed by the boost, are also directly sensitive to the coefficients of the odd parity $\tilde{\kappa}_{0+}$ tensor (see appendix A, and a recent submission that proposes a static electromagnetic test[29]). For the IS experiment analysed in the previous section, the sensitivity to the $\tilde{\kappa}_{0+}$ coefficients is only of parts in $10^9$ or less, and was not presented since cavity resonator experiments have constrained these parameters to parts in $10^{11}$ indirectly through the boost dependence[14,15]. In contrast, the same experiments have measured the parity even parameters directly at parts in $10^{15}$. The boost dependence suppresses the sensitivity by the ratio of the earth's orbital speed to the speed of light, which is of order $10^{-4}$.

In this section we consider the possibility of measuring the odd parity and scalar parameters of the SME with a much higher sensitivity than the IS experiments using a Mach-Zehnder (MZ) interferometer. Interferometers offer the possibility of very sensitive tests at microwave[30,31] and optical frequencies[32]. For this concept to work, each arm of the interferometer must have a different phase dependence on $\tilde{\kappa}_{tr}$ and $\tilde{\kappa}_{0+}$. In this section we show that this is possible if one arm of the interferometer is filled with a magnetic medium. Also, we introduce resonant and power-recycling techniques based on travelling wave resonators, and show that a sensitivity of better than $10^{-15}$ to $\tilde{\kappa}_{0+}$ and $10^{-11}$ to $\tilde{\kappa}_{tr}$ is possible.

The description of electrodynamics in the photon sector of the SME is essentially an extension of Maxwell's equations, where similar SME equations can be written as shown in equation (1). However, the Lagrangian of the SME in the photon sector necessitates that the constitutive relations, given by equation (2), are more general. For the purpose of this work it is sufficient if we assume that the permeability and permittivity are isotropic. This means that the $\vec{\vec{\varepsilon}}_r$ and $\vec{\vec{\mu}}_r$ tensors are diagonal with all three components equal to the scalar permittivity and permeability given by $\varepsilon_r$ and $\mu_r$ respectively. It is implicit in the form of the SME equation (2) that the matter sector does not contribute to the Lorentz violating parameters, as they are not altered in any way by the magnetization or polarization of the material.

### A. Plane wave solution in an isotropic medium

Using the SME equations given in (1) and (2), the simple problem of uniform plane wave propagation in an infinite isotropic medium is considered. Since an interferometer measures phase, the idea here is to calculate the effects of the odd parity and scalar Lorentz violating parameters on the propagation constant of the plane wave in isotropic media in a similar way to (5) in vacuum. In general $\sigma \neq 0$ but for the purposes of the work we assume $\sigma = 0$ (since the magnitude of $\sigma$ as been shown to be less than parts in $10^{32}$)[11,12] and in this case it is not necessary to consider birefringent effects as polarized waves do not change polarization with propagation. From this assumption, (which is equivalent to setting 10 of the 19 possible Lorentz violating parameters to zero) we proceed to calculate the effect of the remaining Lorentz violating parameters on the propagation of plane and guided waves in isotropic media.

First, in the laboratory frame it is assumed that the plane wave is propagating in the $z$ direction at a frequency $\omega$ and propagation constant $\beta$, and thus the electric and magnetic fields have the form;

$$\boldsymbol{E} = \left(E_{xo}\hat{\boldsymbol{x}} + E_{yo}\hat{\boldsymbol{y}} + E_{zo}\hat{\boldsymbol{z}}\right)e^{-i\beta z}e^{i\omega t} \tag{13}$$

$$\boldsymbol{H} = \left(H_{xo}\hat{\boldsymbol{x}} + H_{yo}\hat{\boldsymbol{y}} + H_{zo}\hat{\boldsymbol{z}}\right)e^{-i\beta z}e^{i\omega t} \tag{14}$$

Then, by substituting (13) and (14) into (2) the flux densities $\boldsymbol{D}$ and $\boldsymbol{B}$ may be calculated in terms of the amplitudes of $\boldsymbol{E}$ and $\boldsymbol{H}$. Then by substituting $\boldsymbol{E}$ and $\boldsymbol{B}$ into (1c) and $\boldsymbol{H}$ and $\boldsymbol{D}$ into (1d), to leading order in the Lorentz violating terms, one can show that $E_{zo} \sim H_{zo} \sim 0$ and one is left to solve the following leading-order equation;

$$\begin{bmatrix} -\varepsilon_r - (\kappa_{DE})_{lab}^{xx} & 0 & 0 & \xi - \mu_r(\kappa_{DB})_{lab}^{xy} \\ 0 & -\varepsilon_r - (\kappa_{DE})_{lab}^{yy} & -\xi - \mu_r(\kappa_{DB})_{lab}^{yx} & 0 \\ 0 & \xi - \mu_r(\kappa_{HE})_{lab}^{xy} & \mu_r - \mu_r^2(\kappa_{HB})_{lab}^{xx} & 0 \\ -\xi - \mu_r(\kappa_{HE})_{lab}^{yx} & 0 & 0 & \mu_r - \mu_r^2(\kappa_{HB})_{lab}^{yy} \end{bmatrix} \begin{bmatrix} E_{xo} \\ E_{yo} \\ \sqrt{\dfrac{\mu_o}{\varepsilon_o}} H_{xo} \\ \sqrt{\dfrac{\mu_o}{\varepsilon_o}} H_{yo} \end{bmatrix} = 0 \qquad (15)$$

Here all $\kappa$ matrices are written in the laboratory frame and $\xi = \beta c/\omega$ or $\beta/k_0$, where $k_0 = 2\pi/\lambda_V$, is the propagation constant as calculated in vacuum with no Lorentz violating terms and $\lambda_V$ is the wavelength in vacuum.

Non-trivial solutions are obtained when the determinant of the 4 by 4 matrix of (15) is set to zero. There are four solutions relating to two linear polarizations travelling positively and negatively along the z direction. To leading order, the propagation constants are calculated to be;

$$\frac{\beta_{xy}^{\uparrow}}{k} = 1 + \sqrt{\frac{\mu_r}{\varepsilon_r}}(\kappa_{DB})_{lab}^{xy} + \frac{1}{2\varepsilon_r}(\kappa_{DE})_{lab}^{xx} - \frac{\mu_r}{2}(\kappa_{HB})_{lab}^{yy} \qquad (16a)$$

$$\frac{\beta_{xy}^{\downarrow}}{k} = -1 + \sqrt{\frac{\mu_r}{\varepsilon_r}}(\kappa_{DB})_{lab}^{xy} - \frac{1}{2\varepsilon_r}(\kappa_{DE})_{lab}^{xx} + \frac{\mu_r}{2}(\kappa_{HB})_{lab}^{yy} \qquad (16b)$$

$$\frac{\beta_{yx}^{\uparrow}}{k} = 1 - \sqrt{\frac{\mu_r}{\varepsilon_r}}(\kappa_{DB})_{lab}^{yx} + \frac{1}{2\varepsilon_r}(\kappa_{DE})_{lab}^{yy} - \frac{\mu_r}{2}(\kappa_{HB})_{lab}^{xx} \qquad (16c)$$

$$\frac{\beta_{yx}^{\downarrow}}{k} = -1 - \sqrt{\frac{\mu_r}{\varepsilon_r}}(\kappa_{DB})_{lab}^{yx} - \frac{1}{2\varepsilon_r}(\kappa_{DE})_{lab}^{yy} + \frac{\mu_r}{2}(\kappa_{HB})_{lab}^{xx} \qquad (16d)$$

The superscript refers to the direction of propagation along the z-axis (↑ positive and ↓ negative), the subscripts label the directions of the ***E*** and ***H*** field polarizations respectively, and $k = \sqrt{\varepsilon_r \mu_r}.k_o$ is the propagation constant of the plane wave in the medium. Because we have selected $\sigma = 0$, solutions in the SME are not birefringent, and when we transform (16) to the sun centred frame we recover solutions that are independent of polarization for the leading order $\tilde{\kappa}_{0+}$ and $\tilde{\kappa}_{tr}$ terms (see appendix A). Thus, it is irrelevant what polarization we choose for the following theoretical analysis. Also, any linear combinations of (16) are also solutions of (15), and to maintain consistency with the vacuum equation (5), the following linear combinations of (16) are considered;

$$\frac{\beta^{\uparrow}}{k} = \frac{\beta_{xy}^{\uparrow}}{2k} + \frac{\beta_{yx}^{\uparrow}}{2k} = 1 + \frac{1}{2}\sqrt{\frac{\mu_r}{\varepsilon_r}}\left((\kappa_{DB})_{lab}^{xy} - (\kappa_{DB})_{lab}^{yx}\right) + \frac{1}{4\varepsilon_r}\left((\kappa_{DE})_{lab}^{xx} + (\kappa_{DE})_{lab}^{yy}\right) - \frac{\mu_r}{4}\left((\kappa_{HB})_{lab}^{xx} + (\kappa_{HB})_{lab}^{yy}\right) \qquad (17a)$$

$$\frac{\beta^{\downarrow}}{k} = \frac{\beta_{xy}^{\downarrow}}{2k} + \frac{\beta_{yx}^{\downarrow}}{2k} = -1 + \frac{1}{2}\sqrt{\frac{\mu_r}{\varepsilon_r}}\left((\kappa_{DB})_{lab}^{xy} - (\kappa_{DB})_{lab}^{yx}\right) - \frac{1}{4\varepsilon_r}\left((\kappa_{DE})_{lab}^{xx} + (\kappa_{DE})_{lab}^{yy}\right) + \frac{\mu_r}{4}\left((\kappa_{HB})_{lab}^{xx} + (\kappa_{HB})_{lab}^{yy}\right) \qquad (17b)$$

Equation (17) could have directly been derived from (5) by just making the following substitutions, $k_o \to k$, $\kappa_{DB} \to \sqrt{\frac{\mu_r}{\varepsilon_r}}\kappa_{DB}$, $\kappa_{DE} \to \frac{1}{\varepsilon_r}\kappa_{DE}$ and $\kappa_{HB} \to \mu_r \kappa_{HB}$, and referring the propagation with respect to the laboratory frame. This linear combination describes a wave travelling in the *z* direction with equal amounts of polarization in the *x* and *y* direction.

Now we consider the phase shift recorded by a MZ interferometer with two arms containing different media and with plane waves travelling in the positive z direction as shown in figure 2.

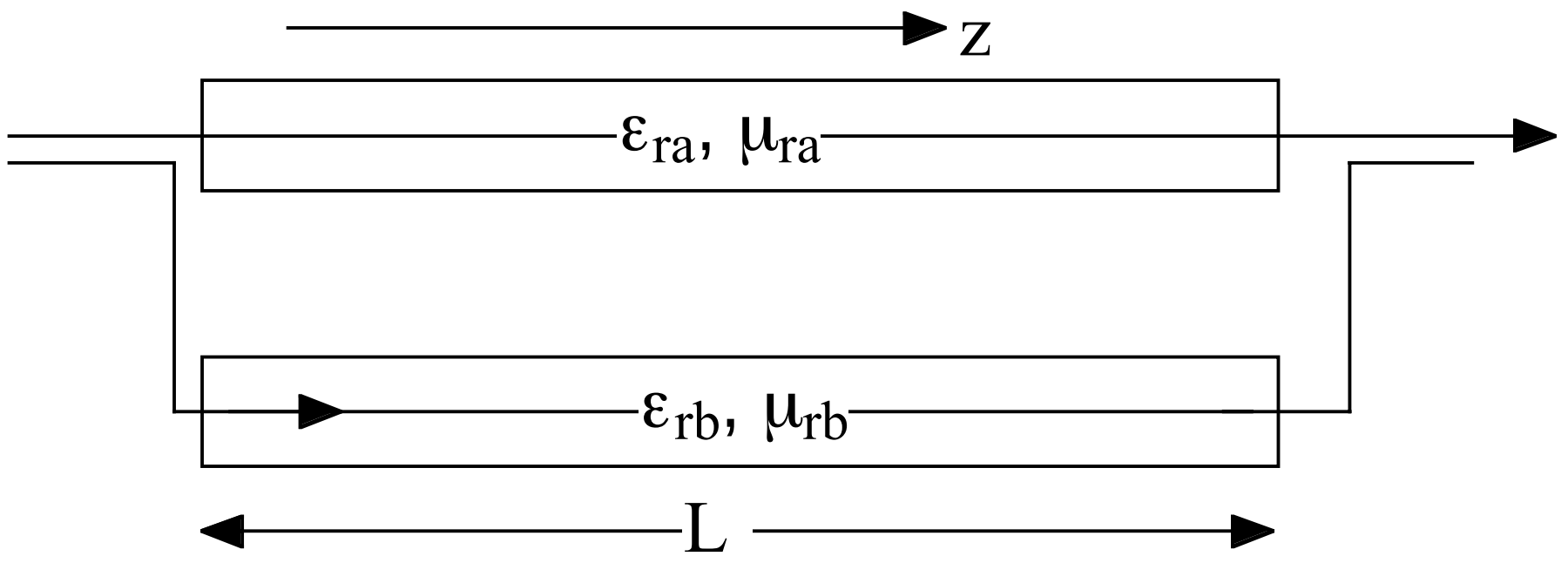

Figure 2. Mach- Zehnder (MZ) interferometer of length *L*, with two arms of permittivity $\varepsilon_{ra}$ and $\varepsilon_{rb}$, and permeability of $\mu_{ra}$ and $\mu_{rb}$ respectively.

For a MZ interferometer of length *L*, the phase shift, $\Delta\theta^{\uparrow}$, at the output is given by the difference of phase gained along the arms *a* and *b*, and is given by;

$$\Delta\theta^{\uparrow} = \theta_a^{\uparrow} - \theta_b^{\uparrow} = L\left(\beta_a^{\uparrow} - \beta_b^{\uparrow}\right) =$$
$$\frac{2\pi L}{\lambda_v}\left[\begin{array}{l}\left(\sqrt{\mu_{ra}\varepsilon_{ra}} - \sqrt{\mu_{rb}\varepsilon_{rb}}\right) + \frac{\left(\mu_{ra} - \mu_{rb}\right)}{2}\left(\left(\kappa_{DB}\right)_{lab}^{xy} - \left(\kappa_{DB}\right)_{lab}^{yx}\right) + \\ \frac{1}{4}\left(\sqrt{\frac{\mu_{ra}}{\varepsilon_{ra}}} - \sqrt{\frac{\mu_{rb}}{\varepsilon_{rb}}}\right)\left(\left(\kappa_{DE}\right)_{lab}^{xx} + \left(\kappa_{DE}\right)_{lab}^{yy}\right) - \frac{1}{4}\left(\sqrt{\mu_{ra}^3\varepsilon_{ra}} - \sqrt{\mu_{rb}^3\varepsilon_{rb}}\right)\left(\left(\kappa_{HB}\right)_{lab}^{xx} + \left(\kappa_{HB}\right)_{lab}^{yy}\right)\end{array}\right] \quad (18)$$

In general there is a phase difference between the two arms, and a modified sensitivity to the Lorentz violating coefficients depending on the permittivity and permeability of each arm. When transformed to the sun centered frame, only the $\left(\kappa_{DB}\right)_{lab}^{jk}$lab terms give rise to non-zero and time varying $\tilde{\kappa}_{0+}$ and $\tilde{\kappa}_{tr}$ terms (i.e. see Appendix A), and from (18) it is evident that the coefficients of $\left(\kappa_{DB}\right)_{lab}^{xy}$ and $\left(\kappa_{DB}\right)_{lab}^{yx}$ only depends on the magnetic properties of the two arms and not the dielectric. The reason is because of two competing effects. First, the sensitivity is reduced by the square root of the permittivity due to the nature of the SME solution from (1) and (2). This is compensated by an increase in phase sensitivity due to the phase length of the interferometer arm being enhanced by the same amount, so the phase shift becomes independent of permittivity. Conversely, for the permeability the two effects combine rather than cancel, so the phase shift becomes proportional to permeability.

The sensitivity of this experiment is proportional to the interferometer arm length divided by the wavelength, which is equal to the number of wavelengths along the arm of the interferometer. Thus, the sensitivity may be increased with long arms and short wavelengths. A further method to improve the sensitivity is through resonant recycling techniques, which have been developed for other applications at both microwave and optical frequencies. These techniques typically improve phase sensitive measurements by up to three orders of magnitude.[30-32] Two different types of recycling are shown in figures 3 and 4. First, we consider the equivalent of resonant cavities in each arm of the interferometer (figure 3), then we consider a power recycling technique (figure 4). In this example we only keep the desired response to the $(\kappa_{DB})_{lab}^{jk}$ coefficients, and set the others to zero.

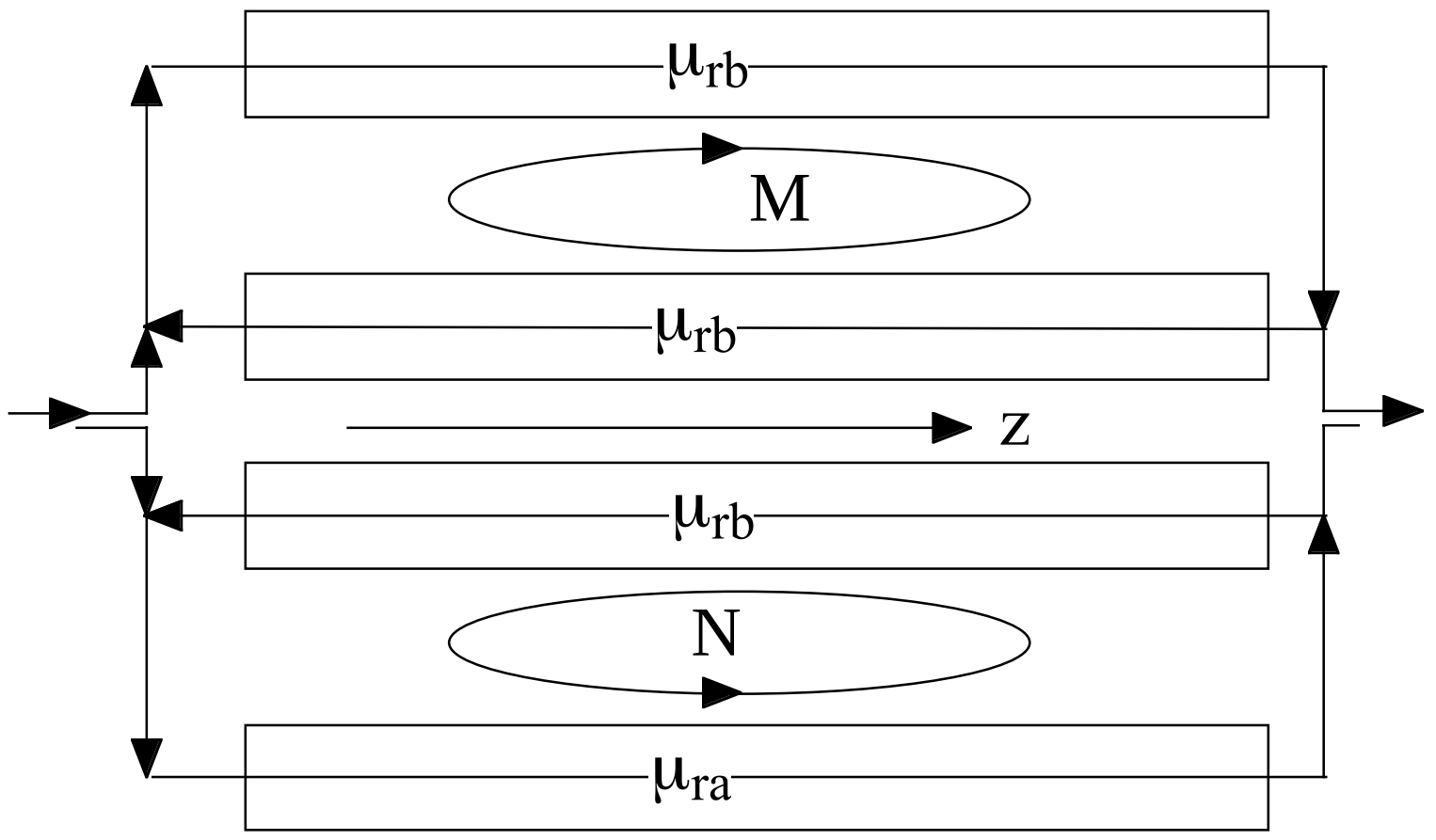

Figure 3. Resonant recycling technique with travelling wave resonators in each arm of the interferometer.

If we define *N* and *M* as the number of recycles in each arm, then the phase shift at the output becomes;

$$\Delta\theta = \left((N+1)\theta_a^{\uparrow} + N\theta_b^{\downarrow}\right) - \left((M+1)\theta_b^{\uparrow} + M\theta_b^{\downarrow}\right) = (N+1)\left(\mu_{ra} - \mu_{rb}\right)\frac{2\pi L}{\lambda_v}\left[\frac{(\kappa_{DB})_{lab}^{xy} - (\kappa_{DB})_{lab}^{yx}}{2}\right] \quad (19)$$

Note, the phase shift turns out to be independent of the number of *M* recycles but proportional to *N*. This is because the arm with *N* recycles has the symmetry broken by the dissimilar forward and return paths, which permits an enhancement in sensitivity directly to the parity-odd coefficients. Conversely, the other arm does not, however, *M* is required to be non-zero so the dispersion in both arms will be similar, and the phase noise of the oscillator driving the interferometer will be suppressed. The parity-odd resonant recycling arm is essentially an asymmetric travelling wave resonator. The same principle may be used to make the frequency of a resonant cavity directly sensitive to the odd parity coefficients, this is illustrated in section IV.

Another way to increase the sensitivity is through power recycling. The equivalent configuration for this type of interferometer is shown in figure 4.

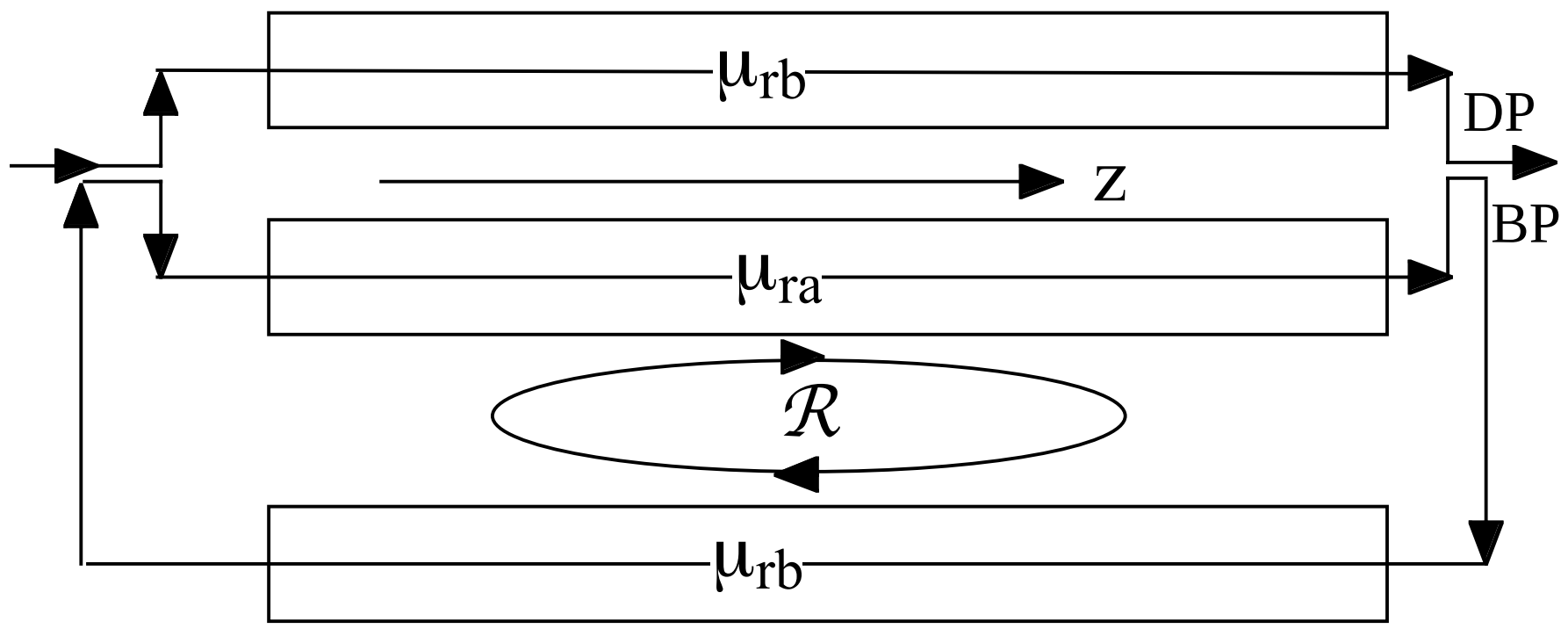

Figure 4. Power-recycling technique where the output power at the Bright Port (BP) of the interferometer is fed back and added to the input. The Dark Port (DP) is the phase sensitive output.

In this case the phase shift at the output becomes;

$$\Delta\theta = (\mathcal{R}+1)\left(\mu_{ra}-\mu_{rb}\right)\frac{2\pi L}{\lambda_v}\left[\frac{\left(\kappa_{DB}\right)_{lab}^{xy}-\left(\kappa_{DB}\right)_{lab}^{yx}}{2}\right] \tag{20}$$

where $\mathcal{R}$ is the power recycling factor, which may be as large as 1000. This is a common technique used to enhance the sensitivity of a gravitational wave interferometer, and it is possible to implement both power and resonant recycling simultaneously.[32]

### B. Perturbation equation for waveguides in the SME

It is more practical to use transmission lines and waveguides to propagate waves in an interferometer at microwave frequencies, while at optical frequencies the plane wave solution may be kept, unless optical fibers are used as a means of transmission. In this section we derive a technique to calculate the effects of the guiding medium in the SME as long as $\sigma = 0$ in (5). The difficulties encountered in solving boundary-valued problems in the SME are greater than with Maxwell's equations or for the leading-order plane wave SME solution given in (16a-d). In this section we aim for a more general solution that implements perturbation analysis so the leading order solution may be calculated from fields derived from Maxwell's equations. This analysis is similar to that of Kostelecky and Mewes, which calculates the leading order perturbation in frequency for a resonant cavity.[11] The analysis starts with the quadratic lemma for two electromagnetic processes ($\boldsymbol{E_0}$, $\boldsymbol{H_0}$, $\boldsymbol{D_0}$, $\boldsymbol{B_0}$) and ($\boldsymbol{E}$, $\boldsymbol{H}$, $\boldsymbol{D}$, $\boldsymbol{B}$) at different oscillation frequencies, $\omega_0$ and $\omega$;

$$\nabla.(\boldsymbol{E}\times\boldsymbol{H}_0^* + \boldsymbol{E}_0^*\times\boldsymbol{H}) + i\omega\boldsymbol{H}_0^*.\boldsymbol{B} - i\omega_0\boldsymbol{B}_0^*.\boldsymbol{H} + i\omega\boldsymbol{E}_0^*.\boldsymbol{D} - i\omega_0\boldsymbol{D}_0^*.\boldsymbol{E} = 0 \tag{21}$$

We base our analysis on the perturbation method, which has been used by Gurevich[33] for computing perturbed propagation constants of modes in waveguides due to a small object. In this case we use a similar technique, but regard the SME fields as perturbations of the Maxwell fields. Here we assume that the fields ($\boldsymbol{E_0}$, $\boldsymbol{H_0}$, $\boldsymbol{D_0}$, $\boldsymbol{B_0}$) are the Maxwell fields given by (2) when the $\kappa$ matrices are set to zero, and that ($\boldsymbol{E}$, $\boldsymbol{H}$, $\boldsymbol{D}$, $\boldsymbol{B}$) are the SME fields given by (2) when the $\kappa$ matrices are non-zero. Thus, by substituting (2) into (21) we can derive the fundamental Lemma of the SME in the photon sector;

$$\begin{aligned}\nabla.(\boldsymbol{E}\times\boldsymbol{H}_0^* + \boldsymbol{E}_0^*\times\boldsymbol{H}) = {} & i\frac{\omega_0}{\mu_0}\boldsymbol{B}_0^*.\vec{\mu}_r^{-1}.\boldsymbol{B} - i\frac{\omega}{\mu_0}\left(\overleftrightarrow{\mu}_r^{-1*}.\boldsymbol{B}_0^*\right).\boldsymbol{B} + i\omega_0\varepsilon_0\left(\overleftrightarrow{\varepsilon}_r^*.\boldsymbol{E}_0^*\right).\boldsymbol{E} - i\omega\varepsilon_0\boldsymbol{E}_0^*.\overleftrightarrow{\varepsilon}_r.\boldsymbol{E} + \\ & i\omega_0\left(\frac{1}{\mu_0}\boldsymbol{B}_0^*.\kappa_{HB}.\boldsymbol{B} + \sqrt{\frac{\varepsilon_0}{\mu_0}}\boldsymbol{B}_0^*.\kappa_{HE}.\boldsymbol{E}\right) - i\omega\left(\varepsilon_0\boldsymbol{E}_0^*.\kappa_{DE}.\boldsymbol{E} + \sqrt{\frac{\varepsilon_0}{\mu_0}}\boldsymbol{E}_0^*.\kappa_{DB}.\boldsymbol{B}\right)\end{aligned} \tag{22}$$

This Lemma may be used to calculate the frequency perturbation in the SME of a resonator.[11] For our case we use the Lemma to calculate the perturbation of the propagation constant for a guided wave. Here, we assume the frequency in the SME is the same for a propagating mode as calculated by Maxwell's equations, so that $\omega = \omega_0$. Finally, if the permittivity and permeability tensors are Hermitian (as is the case for isotropic to gyrotropic media) then the first four terms on the right-hand side of equation (22) cancel, and it becomes;

$$\nabla.(\boldsymbol{E}\times\boldsymbol{H}_0^* + \boldsymbol{E}_0^*\times\boldsymbol{H}) = i\omega_0\left(-\varepsilon_0\boldsymbol{E}_0^*.\kappa_{DE}.\boldsymbol{E} + \frac{1}{\mu_0}\boldsymbol{B}_0^*.\kappa_{HB}.\boldsymbol{B} - \sqrt{\frac{\varepsilon_0}{\mu_0}}\boldsymbol{E}_0^*.\kappa_{DB}.\boldsymbol{B} + \sqrt{\frac{\varepsilon_0}{\mu_0}}\boldsymbol{B}_0^*.\kappa_{HE}.\boldsymbol{E}\right) \tag{23}$$

The solutions will be travelling waves so we may assume they are of the form;

$$\boldsymbol{E} = \boldsymbol{E}_c e^{-i\beta z}, \boldsymbol{H} = \boldsymbol{H}_c e^{-i\beta z}, \boldsymbol{E}_0 = \boldsymbol{E}_{0c} e^{-i\beta_0 z}, \boldsymbol{H}_0 = \boldsymbol{H}_{0c} e^{-i\beta_0 z} \tag{24}$$

Here $\beta_0$ is the propagation constant calculated by Maxwell's equations, $\beta$ is the propagation constant in the SME, ($\boldsymbol{E}_{0c}$, $\boldsymbol{H}_{0c}$, $\boldsymbol{D}_{0c}$, $\boldsymbol{B}_{0c}$) and ($\boldsymbol{E}_c$, $\boldsymbol{H}_c$, $\boldsymbol{D}_c$, $\boldsymbol{B}_c$) are the respective vector phasor amplitudes of the electromagnetic field for Maxwell's equations and in the SME. Substituting (24) into (23) and then integrating over the surface shown in figure 5, (23) becomes;

$$\begin{aligned}&\int_L (\boldsymbol{E}_c \times \boldsymbol{H}_{0c}^* + \boldsymbol{E}_{0c}^* \times \boldsymbol{H}_c).\boldsymbol{n}_0 dL - i(\beta - \beta_0)\int_S \left(\boldsymbol{E}_c \times \boldsymbol{H}_{0c}^* + \boldsymbol{E}_{0c}^* \times \boldsymbol{H}_c\right).\boldsymbol{z}_0 dS \\ &= i\omega_0 \int_S \left( -\varepsilon_0 \boldsymbol{E}_{0c}^*.\kappa_{DE}.\boldsymbol{E}_c + \frac{1}{\mu_0}\boldsymbol{B}_{0c}^*.\kappa_{HB}.\boldsymbol{B}_c - \sqrt{\frac{\varepsilon_0}{\mu_0}}\boldsymbol{E}_{0c}^*.\kappa_{DB}.\boldsymbol{B}_c + \sqrt{\frac{\varepsilon_0}{\mu_0}}\boldsymbol{B}_{0c}^*.\kappa_{HE}.\boldsymbol{E}_c \right) dS\end{aligned} \tag{25}$$

Here $\boldsymbol{n}_0$ is a unit vector along the normal to the curve $\mathcal{L}$, which lies in the plane of the waveguide cross section and $\boldsymbol{z}_0$ is a unit vector along the waveguide axis.

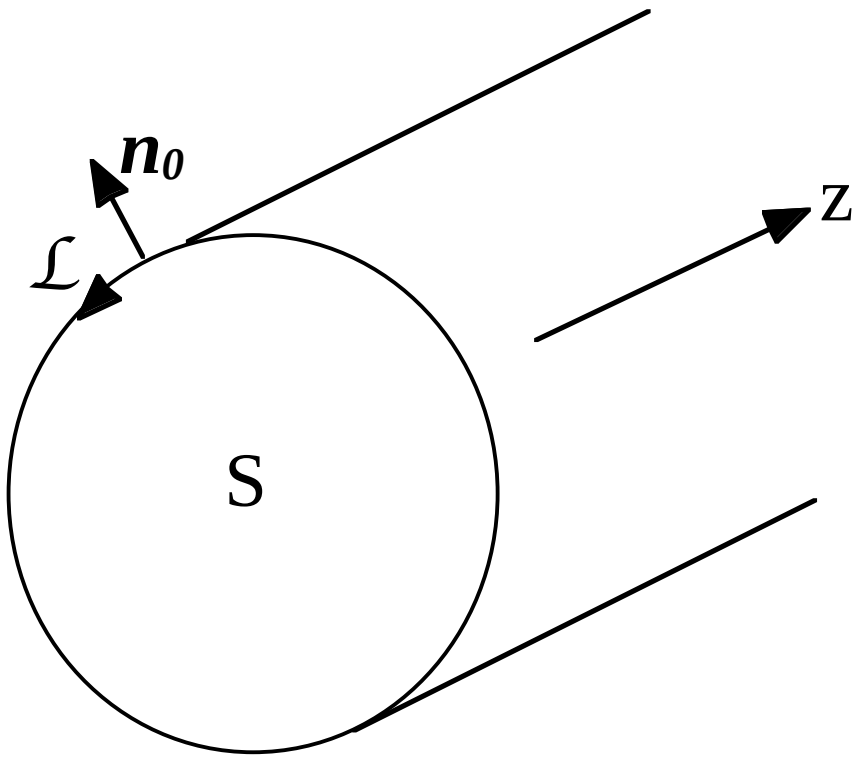

Figure 5. Cross-section of the transmission medium

Assuming no Lorentz violation in the matter sector, the boundary conditions of the tangential electric field in the photon sector of the SME have been shown to be equal to zero, equivalent to the Maxwell case.[11] Thus, the first term in (25) is zero, and to calculate the leading order effects on the propagation constant we replace the perturbed SME fields with the unperturbed Maxwell fields and equate imaginary parts on both sides of equation (25), and applying the relationship $\kappa_{DB} = -\kappa_{HE}^T$, to give;

$$\beta - \beta_0 = ck_0 \frac{\int_S \left( \varepsilon_0 \boldsymbol{E}_{0c}^*.\kappa_{DE}.\boldsymbol{E}_{0c} - \frac{1}{\mu_0}\boldsymbol{B}_{0c}^*.\kappa_{HB}.\boldsymbol{B}_{0c} + 2\,\mathrm{Re}\left[\sqrt{\frac{\varepsilon_0}{\mu_0}}\boldsymbol{E}_{0c}^*.\kappa_{DB}.\boldsymbol{B}_{0c}\right]\right) dS}{\int_S \left(\boldsymbol{E}_{0c} \times \boldsymbol{H}_{0c}^* + \boldsymbol{E}_{0c}^* \times \boldsymbol{H}_{0c}\right).\boldsymbol{z}_0 dS} \tag{26}$$

However, the observable for a MZ interferometer is the phase difference between the two arms $\Delta\theta$. Thus, in the laboratory frame we may rewrite (26) in terms of phase shift induced by Lorentz violation for a single mode propagating along a single waveguide as;

$$\theta = (\beta - \beta_0)L = (\mathcal{M}_{DE})_{lab}^{jk}(\kappa_{DE})_{lab}^{jk} + (\mathcal{M}_{HB})_{lab}^{jk}(\kappa_{HB})_{lab}^{jk} + (\mathcal{M}_{DB})_{lab}^{jk}(\kappa_{DB})_{lab}^{jk} \tag{27}$$

where L is the length of propagation along the z-axis and,

$$(\mathcal{M}_{DE})_{lab}^{jk} = \frac{2\pi L}{\lambda_v} \frac{\int_S \left(\sqrt{\frac{\varepsilon_0}{\mu_0}} \boldsymbol{E}_{0c}^{j*}.\boldsymbol{E}_{0c}^{k}\right) dS}{\int_S \left(\boldsymbol{E}_{0c} \times \boldsymbol{H}_{0c}^* + \boldsymbol{E}_{0c}^* \times \boldsymbol{H}_{0c}\right).\boldsymbol{z}_0 dS} \tag{28a}$$

$$\left(\mathcal{M}_{HB}\right)_{lab}^{jk} = \frac{2\pi L}{\lambda_v}\frac{-\int_S\left(\frac{1}{\sqrt{\varepsilon_0\mu_0^3}}B_{0c}^{j*}.B_{0c}^{k}\right)dS}{\int_S\left(\boldsymbol{E}_{0c}\times\boldsymbol{H}_{0c}^{*}+\boldsymbol{E}_{0c}^{*}\times\boldsymbol{H}_{0c}\right).\boldsymbol{z}_0 dS} \tag{28b}$$

$$\left(\mathcal{M}_{DB}\right)_{lab}^{jk} = \frac{2\pi L}{\lambda_v}\frac{2\,\mathrm{Re}\left[\int_S\left(\frac{1}{\mu_0}\boldsymbol{E}_{0c}^{j*}.B_{0c}^{k}\right)dS\right]}{\int_S\left(\boldsymbol{E}_{0c}\times\boldsymbol{H}_{0c}^{*}+\boldsymbol{E}_{0c}^{*}\times\boldsymbol{H}_{0c}\right).\boldsymbol{z}_0 dS} \tag{28c}$$

Thus, if the Maxwell fields are known for the propagating solution, the phase shift due to a leading order Lorentz violating terms in the SME may be calculated. It is easily shown that equation (27) is consistent with the plane wave solution calculated previously.

For the general guided wave, the sensitivity to odd parity and scalar coefficients is proportional to the difference between the two coefficients $\left(\mathcal{M}_{DB}\right)_{lab}^{jk} - \left(\mathcal{M}_{DB}\right)_{lab}^{kj}$ (j≠k), which is $\frac{2\pi L}{\lambda_v}\mu_r$ for the propagating plane wave independent of polarization. We have also undertaken this calculation for many different modes in many types of waveguides (but not presented here), including cylindrical, coaxial and helical guiding structures. In all cases the $\left(\mathcal{M}_{DB}\right)_{lab}^{jk} - \left(\mathcal{M}_{DB}\right)_{lab}^{kj}$ coefficient was the same as the plane wave case (for example, the calculation for TE modes in a rectangular waveguide is given in Appendix B). For the interferometer experiment the observable is phase difference between two guided waves. The equivalent $\mathcal{M}$ matrices for this case are then given by the difference of the two arms, which will be non-zero when the permeability is different.

In general the wave need not be propagating along the *z* direction of the laboratory frame. To calculate the $\mathcal{M}$ tensors for a wave propagating along another arbitrary direction, the tensor must undergo a rotation. To detect the signal, we rely on the time dependence to modulate the phase, which can be determined in the standard way as described in Kostelecky and Mewes[11] by transforming from the lab (rotating or non-rotating) to the sun centred celestial frame, as given in appendix A.

### C. Proposed interferometer experiment

Sensitive interferometers have been developed at optical[32] and microwave frequencies.[31, 34] However this experiment requires the availability of low loss magnetic material at the respective frequency. Low loss materials are only possible at microwave frequencies with a relative permeability less than unity above the magnetic spin resonance of the material. For example, one could use YIG (Yttrium Iron Garnet), which has a permeability of about 0.9 and loss tangent of $10^{-4}$ at 10 GHz[35]. In this paper we have chosen YIG for our example experiment. Alternatively, at optical frequencies magnetic effects may be induced using magnetic polaritons,[36-38] and if the ratio of length to wavelength can be increased a more sensitive measurement may be viable. The sensitivity of this type of experiment will be dependent on a host of other technical factors, like the available power from the frequency source, the phase balance between the interferometer arms etc. These dependencies have been largely discussed in the papers that implement low noise interferometers, and we leave out the details here. For a thermal noise limited microwave interferometer with no recycling, the typical square root spectral density of phase noise $\sqrt{S_\phi}$ is conservatively of order $10^{-9}$ rads/√Hz.[31] Chopping techniques to lock the interferometer output to the DP (null measurement) can further reduce the flicker noise corner,[32] and rotation of the experiment can ensure that the thermal noise limit is reached. In this case the sensitivity of the phase measurement is dependent on the observation time, $\tau_{obs}$, and given by

$$\delta\theta\big|_{SNR=1} = \sqrt{\frac{S_\phi}{N_c \tau_{obs}}} \quad . \qquad (29)$$

where $N_c$ is the number (or fraction) of cycles in one second. If the interferometer arms labelled *b*, as shown in figure 2 to 4 contain vacuum, then the signal due to the Lorentz violation with power and resonant recycling will be

$$\Delta\theta = -(\mathcal{R}+1)(N+1)\frac{2\pi L}{\lambda_v}(1-\mu_{ra})\left[\frac{(\kappa_{DB})^{xy}_{lab} - (\kappa_{DB})^{yx}_{lab}}{2}\right] \quad . \qquad (30)$$

This means the sensitivity of the measurement may be given with a Signal to Noise Ratio (SNR) of one (or significance of one standard deviation) when

$$\left[\frac{(\kappa_{DB})^{xy}_{lab} - (\kappa_{DB})^{yx}_{lab}}{2}\right]\Bigg|_{SNR=1} = -\frac{\lambda_v}{2\pi L(R+1)(N+1)(1-\mu_{ra})}\sqrt{\frac{S_\phi}{N_c \tau_{obs}}} \quad . \qquad (31)$$

For a rotating 10 GHz interferometer of order one meter long with a recycling factor of $N$ +1= 100 and $\mathcal{R}$ +1 = 100, the estimated sensitivity of (31) is of the order $2.10^{-14}/\sqrt{\tau_{obs}}$ for $N_c$ = 0.05 (20 second rotation period). Thus, a sensitivity of order $10^{-15}$ is possible with only 450 seconds of data. For a non-rotating experiment $N_c = 1.157\times10^{-5}$ (one day rotation period), and, a sensitivity of order $10^{-15}$ is possible with 22.5 days of data. This translates to sensitivity to $\tilde{\kappa}_{tr}$ of order $10^{-11}$ and $\tilde{\kappa}_{0+}$ of order $10^{-15}$. A further benefit from rotation over non-rotating experiments is direct sensitivity to all three independent coefficients of $\tilde{\kappa}_{0+}$ since non-rotating experiments only allow two of the three coefficients to be tested directly (see Appendix A).

## V. POSSIBLE RESONATOR EXPERIMENTS

The fact that resonant and power recycling enhances the phase sensitivity to Lorentz violations suggests that the frequency of an asymmetric travelling wave resonator will also be sensitive. A generic resonator of this type is shown in figure 6.

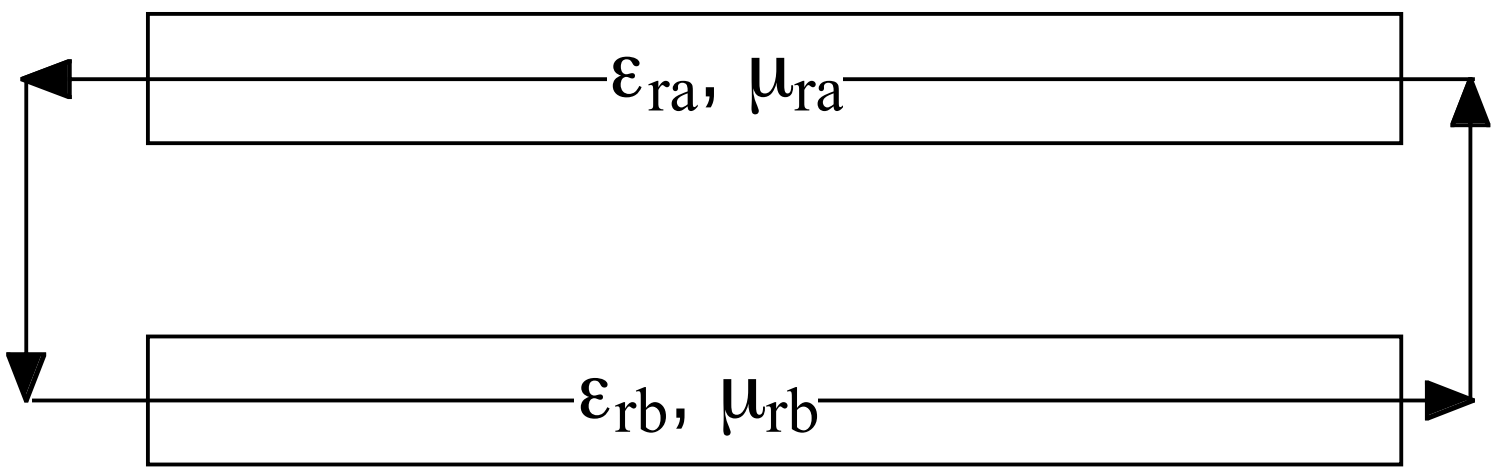

Figure 6. Schematic of a travelling wave resonator with electromagnetic asymmetry on the forward and reverse path of the resonator.

To calculate the frequency shift of such a resonator we use the same equation as derived by Kostelecky and Mewes[11]

$$\frac{\delta\nu}{\nu} = -\frac{1}{4\langle U\rangle}\int_V\left(\varepsilon_0\boldsymbol{E}^*_{0c}.\kappa_{DE}.\boldsymbol{E}_{0c} - \frac{1}{\mu_0}\boldsymbol{B}^*_{0c}.\kappa_{HB}.\boldsymbol{B}_{0c} + 2\,\mathrm{Re}\left[\sqrt{\frac{\varepsilon_0}{\mu_0}}\boldsymbol{E}^*_{0c}.\kappa_{DB}.\boldsymbol{B}_{0c}\right]\right)dV \tag{32}$$

$$\langle U\rangle = \frac{1}{4}\int_V\left(\boldsymbol{E}_{0c}.\boldsymbol{D}^*_{0c} + \boldsymbol{B}_{0c}.\boldsymbol{H}^*_{0c}\right)dV.$$

Thus, for the observable, $\delta\nu/\nu$, the $\mathcal{M}$ matrices become;

$$\left(\mathcal{M}_{DE}\right)^{jk}_{lab} = -\frac{\varepsilon_0}{4\langle U\rangle}\int_V\left(\boldsymbol{E}^{j*}_{0c}.\boldsymbol{E}^{k}_{0c}\right)dV \tag{33a}$$

$$\left(\mathcal{M}_{HB}\right)^{jk}_{lab} = \frac{1}{4\langle U\rangle\mu_0}\int_V\left(\boldsymbol{B}^{j*}_{0c}.\boldsymbol{B}^{k}_{0c}\right)dV \tag{33b}$$

$$\left(\mathcal{M}_{DB}\right)^{jk}_{lab} = -\frac{1}{2\langle U\rangle}\sqrt{\frac{\varepsilon_0}{\mu_0}}\int_V\left(\boldsymbol{E}^{j*}_{0c}.\boldsymbol{B}^{k}_{0c}\right)dV \tag{33c}$$

In this example we only need to consider the coefficients of the $\left(\kappa_{DB}\right)^{jk}_{lab}$, which give rise to non-zero and time varying $\tilde{\kappa}_{0+}$ and $\tilde{\kappa}_{tr}$. Here we assume the wave travels along a waveguide and is returned along a second piece of waveguide of different cross-section, permittivity and permeability (but same length $L$) as shown in figure 6. The integrals over the forward and reverse paths must be undertaken separately, and the fraction of energy in the forward and return path is in general not equal. Without Lorentz violation, resonance occurs when an integer number of half wavelengths fit into $2L$, so that the wave undergoes constructive interference. In practice, such a loop will contain isolators, couplers all contributing to the frequency shift in the SME. However, if long enough most of the energy will be in the two pieces of waveguide, and we can ignore any other component in this example. Also, we assume that the wave propagates as a plane wave along the $z$ direction, with equal amounts of polarization in the $x$ and $y$ direction. In this case the non-zero components of (33c) are;

$$\left(\mathcal{M}_{DB}\right)^{xy}_{lab} - \left(\mathcal{M}_{DB}\right)^{yx}_{lab} = -\frac{\sqrt{\mu_{ra}\varepsilon_{ra}} - \sqrt{\mu_{rb}\varepsilon_{rb}}}{2\left(\varepsilon_{ra}+\varepsilon_{rb}\right)} \tag{34}$$

Hence if the phase velocity, $v_{ph} = c/(\varepsilon_r.\mu_r)^{1/2}$, in medium $a$ is different to medium $b$, the resonator is sensitive to odd parity coefficients, i.e. the frequency shift will be sensitive to $\left(\kappa_{DB}\right)^{jk}_{lab}$ coefficients. Unlike the interferometer, the experiment does not necessarily require magnetic materials, as the loop could be made of a free space and dielectric arm for (34) to be non-zero. This is due to the different nature of the frequency and phase observables. For the MZ interferometer the effect of the dielectric constant was cancelled by the effective increase in the electrical phase length of the interferometer arm (see equation (18)). This is not the case for the resonator in figure 6 when frequency shift is the observable, as any putative frequency shift due to Lorentz violation is not directly dependent on the length.

Next, we calculate the $\mathcal{M}_{DB}$ matrix for arms $a$ and $b$ made from rectangle waveguide and operating in TE mode. Substituting the fields from appendix B, and integrating over the two paths gives the following.

$$\left(\mathcal{M}_{DB}\right)^{xy}_{lab} - \left(\mathcal{M}_{DB}\right)^{yx}_{lab} = -\left(\frac{1}{1+\langle U_b\rangle / \langle U_a\rangle}\left(\frac{c}{v_{pha}\varepsilon_{ra}}\right) - \frac{1}{1+\langle U_a\rangle / \langle U_b\rangle}\left(\frac{c}{v_{phb}\varepsilon_{rb}}\right)\right) \tag{35}$$

$$\text{where } \langle U_a\rangle / \langle U_b\rangle = \left(\frac{\varepsilon_{ra}}{\varepsilon_{rb}}\right)\left(\frac{A_a}{A_b}\right)\left(\frac{1-\left(\frac{c}{v_{phb}}\right)^2}{1-\left(\frac{c}{v_{pha}}\right)^2}\right),$$

$\langle U\rangle = \langle U_a\rangle + \langle U_b\rangle$, and $\langle U_a\rangle$ and $\langle U_b\rangle$ are the average energies in arm *a* and *b* of the resonator respectively. Also, $v_{pha}$ and $v_{phb}$ are the magnitude of the phase velocities from $z$ = 0 to $L$ along arm *a* and $z$ = $L$ to 0 along arm *b* respectively, and $A_a$ and $A_b$ are the areas of cross section of the two waveguides. Equation (35) is a more general solution of (34), and in the limit that both cross-sections approach infinity, we regain the plane wave solution of (34), independent of polarization.

Now we study the case when arm *a* is a low loss dielectric and arm *b* is vacuum, then

$$\left(\mathcal{M}_{DB}\right)^{xy}_{lab} - \left(\mathcal{M}_{DB}\right)^{yx}_{lab} = -\frac{\sqrt{\varepsilon_{ra}}-1}{2\left(\varepsilon_{ra}+1\right)} \quad . \tag{36}$$

A plot of this function is shown in figure 7,

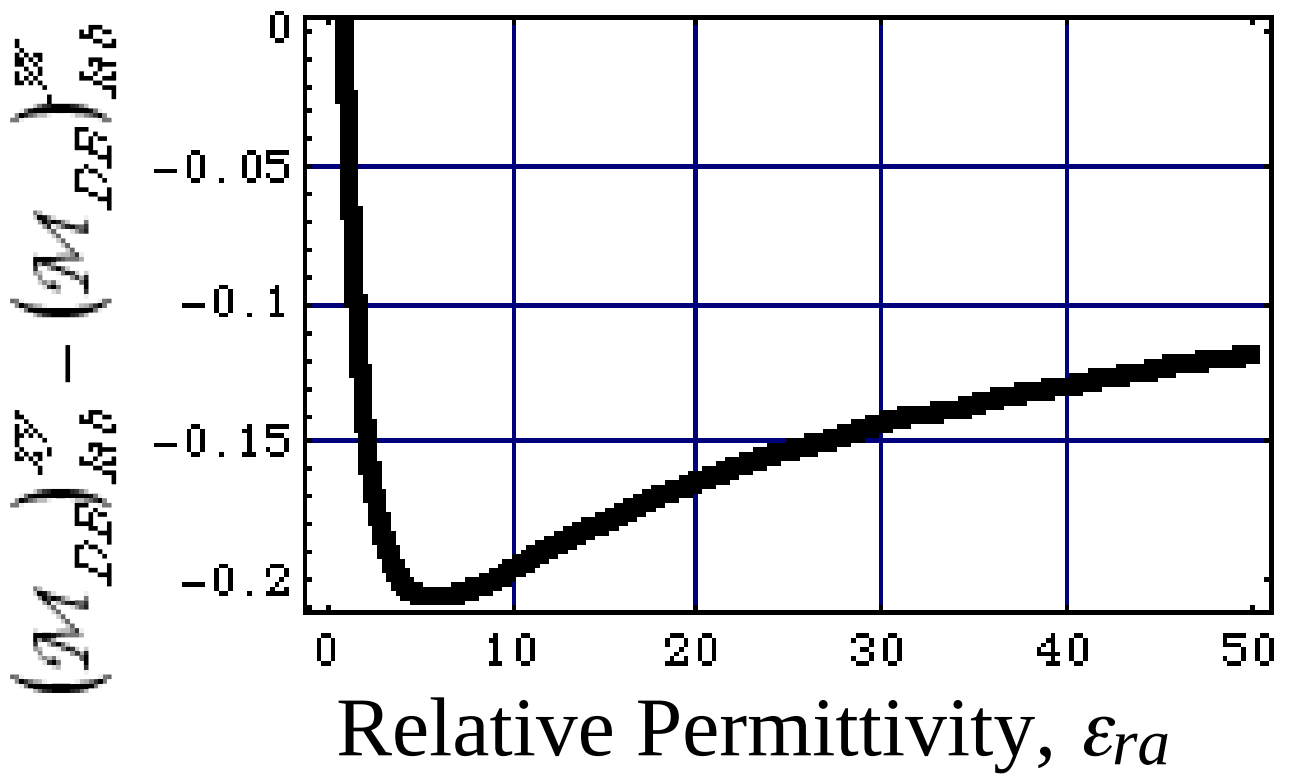

Figure 7. Sensitivity coefficient versus relative permittivity of equation (36).

In this case there is a maximum value of the magnitude, when $\varepsilon_{ra}$ = 5.83 and the coefficient is equal to -0.207. Sapphire with the propagation direction along the c-axis of the crystal is a good choice, because it has a lower loss tangent at microwave frequencies than any other material. For this configuration the mode only samples the perpendicular permittivity of sapphire (~ 9.3), and in this case the magnitude of the coefficient is only slightly smaller than the maximum and equal to 0.20.

In principle (35) may still give a non-zero result when the forward and return paths of the resonator are the same material, but have different phase velocities. This in general can be achieved if the two halves have different cross-sectional dimensions. For this idea to be transformed to a sensitive experiment, a high-Q travelling wave resonator that exhibits asymmetry (as described above) must be developed. High-Q sapphire Whispering Gallery (WG) resonators have been used to do the best Lorentz invariance tests of the polarization independent coefficients to date,[15,17,39] and it is possible to excite them as travelling waves.[41] One way to make a resonator sensitive to this type of measurement is to excite WG travelling wave modes in an asymmetric resonator; some possible configurations are shown in figure 8.

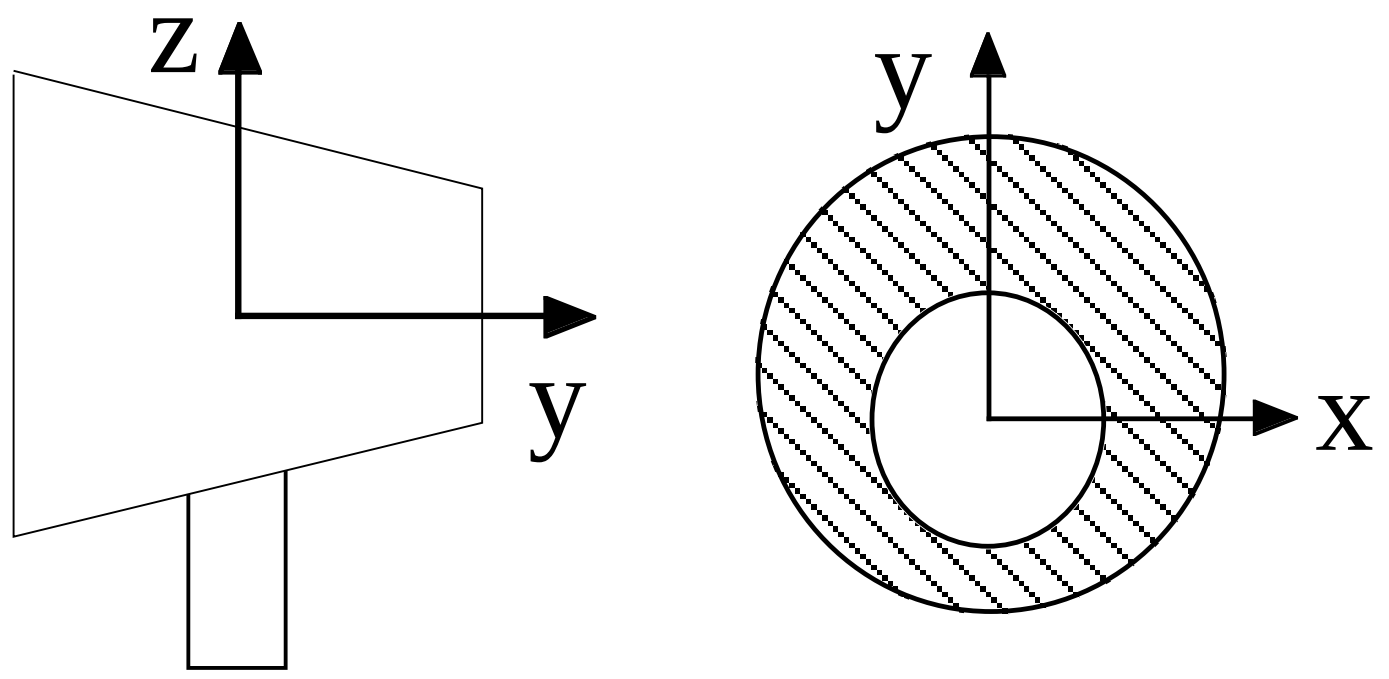

Figure 8. Two possible resonant structures, which exhibit asymmetry around the x-axis of a cylinder. The mode of propagation would be a travelling WG mode around the perimeter of the cylinder[41].

The only way to accurately calculate the sensitivity of such experiments is through numerical simulation, and we will pursue this path in the future. Another possibility is to dope the sapphire crystal with a paramagnetic impurity by an uneven amount along the y-axis of the resonator. Paramagnetic impurities, such as $Cr^{3+}$ add an extra susceptibility due to the Electron Spin Resonance (ESR).[42] In this case the travelling wave would experience a different permeability around the path of resonance, which would break the symmetry. These experiments will also benefit from a further increase in sensitivity with rotation, as has been proposed for the standard cavity experiments.[18]

## VI. CONCLUSION

In the photon sector of the SME numerous experiments have already set limits on 17 of the 19 possible Lorentz violating coefficients. Of these, the 10 coefficients that depend on the polarization have had upper limits set at parts in $10^{32}$ by astrophysical tests.[11,12] While cavity experiments[13-15] have set upper limits on 4 of the 5 polarization independent even parity coefficients at parts in $10^{15}$, and the 3 polarization independent odd parity coefficients at parts in $10^{11}$ (with boost suppression). This work has focused on improving the limits of the odd parity coefficients by investigating experiments that have direct sensitivity and are not suppressed by the boost dependence. Also, we have shown that the same experiments allow the first upper limit of the scalar coefficient to be determined through the boost dependence. Of the experiments undertaken to date, we have shown that IS experiments have the required properties, and by analysing the best experiment[22] we have provided a first upper limit of parts in $10^5$ of the scalar coefficient. Furthermore, we have shown that a magnetically asymmetric Mach-Zehnder interferometer (microwave or optical) may provide a null experiment that is sensitive to the same SME parameters as the IS experiment. We have proposed recycling techniques to further enhance the sensitivity and have shown that the respective sensitivity to the odd parity and scalar coefficients are possible at parts in $10^{15}$ and $10^{11}$ with current technology. The ideas were then extended to asymmetric resonant structures and possible resonator designs were proposed. Future work will concentrate on studying the detailed experimental feasabilities of the interferometer and resonator proposals, with the aim of realising such an experiment within the next years.

**Acknowledgments**

The authors would like to thank Alan Kostelecky and Mathew Mewes for useful discussions. This work was supported by the Australian Research Council.

## APPENDIX A: TRANSFORMATION OF PARITY-ODD AND SCALAR COEFFICIENTS TO THE SUN CENTRED CELESTIAL EQUATORIAL FRAME

In this appendix we undertake to calculate the time dependence of a general experiment sensitive to $\tilde{\kappa}_{tr}$ and $\tilde{\kappa}_{0+}$coefficients with respect to the inertial sun centred celestial frame, in which the Lorentz violating coefficients are constant. To do this we set all terms with respect to the sun centred celestial frame to zero except for the $\tilde{\kappa}_{tr}$ and $\tilde{\kappa}_{0+}$terms, as stringent limits by astrophysical and cavity experiments have already been set. We transform the laboratory kappa tensors to the sun centred frame using rotations $R$ and boosts $v$ as given in.[11]

$$\left(\kappa_{DE}\right)_{lab}^{jk} = T_0^{jkJK}\kappa_{DE}^{JK} - T_1^{kjJK}\kappa_{DB}^{JK} - T_1^{jkJK}\kappa_{DB}^{JK} \quad \text{(A1)}$$

$$\left(\kappa_{HB}\right)_{lab}^{jk} = T_0^{jkJK}\kappa_{HB}^{JK} - T_1^{kjKJ}\kappa_{DB}^{JK} - T_1^{jkKJ}\kappa_{DB}^{JK} \quad \text{(A2)}$$

$$\left(\kappa_{DB}\right)_{lab}^{jk} = T_0^{jkJK}\kappa_{DB}^{JK} + T_1^{kjJK}\kappa_{DE}^{JK} + T_1^{jkJK}\kappa_{HB}^{JK} \quad \text{(A3)}$$

where $T_0^{jkJK} = R^{jJ}R^{kK}$ and $T_1^{jkJK} = R^{jP}R^{kJ}\varepsilon^{KPQ}\,\mathrm{v}^Q/c$. Here, we only list $\tilde{\kappa}_{0+}$coefficients that are directly sensitive to the measurement and not suppressed by the boost term, and we do not list the constant terms only the time varying ones. Setting all components to zero except for $\tilde{\kappa}_{tr}$ and $\tilde{\kappa}_{0+}$ means that the time varying parts of (A1) and (A2) become zero. Then we are just left with the time dependence of (A3), which will modulate the laboratory observable at specific frequencies relating to both the boosts and rotations and we calculate these modulations for stationary and rotating experiments in the following sections. Also, it is noted that all modulations occur around the spin and earth's rotation frequency, and that modulations around twice these values are zero. This is not the case for the established resonator experiments that have put limits on the other parameters.[15, 39]

### 1. Stationary laboratory experiments

Substituting (A3) into (27) or (32) (depending on the observable), the non-zero coefficients of modulation are given in tables 1 to 3. Note that there are no modulation frequencies that directly vary the $\tilde{\kappa}_{0+}^{XY}$ coefficient. This is because the orientation in the sun-centred frame varies only with the rotation of the Earth, and it is this symmetry with respect to the rotation axis, which makes these experiments insensitive to $\tilde{\kappa}_{0+}^{XY}$.

**Table I, Sensitivity coefficients of $\left(\mathrm{v}_\oplus/c\right)\tilde{\kappa}_{tr}$ at the relevant modulation frequencies**

| Modulation | $\left(\mathrm{v}_\oplus/c\right)\tilde{\kappa}_{tr}$ Coefficient |
|---|---|
| $Cos[\omega_\oplus T_\oplus + \Omega_\oplus T]$ | $-(1-Cos[\eta])\left(\left(\mathcal{M}_{DB}\right)_{lab}^{xz} - \left(\mathcal{M}_{DB}\right)_{lab}^{zx}\right)$ |
| $Sin[\omega_\oplus T_\oplus + \Omega_\oplus T]$ | $(1-Cos[\eta])\left(Cos[\chi]\left(\left(\mathcal{M}_{DB}\right)_{lab}^{yz} - \left(\mathcal{M}_{DB}\right)_{lab}^{zy}\right) + Sin[\chi]\left(\left(\mathcal{M}_{DB}\right)_{lab}^{xy} - \left(\mathcal{M}_{DB}\right)_{lab}^{yx}\right)\right)$ |
| $Cos[\omega_\oplus T_\oplus - \Omega_\oplus T]$ | $(1+Cos[\eta])\left(\left(\mathcal{M}_{DB}\right)_{lab}^{xz} - \left(\mathcal{M}_{DB}\right)_{lab}^{zx}\right)$ |
| $Sin[\omega_\oplus T_\oplus - \Omega_\oplus T]$ | $-(1+Cos[\eta])\left(Cos[\chi]\left((\mathcal{M}_{DB})^{yz}_{lab} - (\mathcal{M}_{DB})^{zy}_{lab}\right) + Sin[\chi]\left((\mathcal{M}_{DB})^{xy}_{lab} - (\mathcal{M}_{DB})^{yx}_{lab}\right)\right)$ |

**Table II, Sensitivity coefficients of $\tilde{\kappa}^{XZ}_{0+}$ at the relevant modulation frequencies**

| Modulation | $\tilde{\kappa}^{XZ}_{0+}$ Coefficient |
|---|---|
| $Cos[\omega_\oplus T_\oplus]$ | $\left((\mathcal{M}_{DB})^{xz}_{lab} - (\mathcal{M}_{DB})^{zx}_{lab}\right)$ |
| $Sin[\omega_\oplus T_\oplus]$ | $-Cos[\chi]\left((\mathcal{M}_{DB})^{yz}_{lab} - (\mathcal{M}_{DB})^{zy}_{lab}\right) - Sin[\chi]\left((\mathcal{M}_{DB})^{xy}_{lab} - (\mathcal{M}_{DB})^{yx}_{lab}\right)$ |

**Table III, Sensitivity coefficients of $\tilde{\kappa}^{YZ}_{0+}$ at the relevant modulation frequencies**

| Modulation | $\tilde{\kappa}^{YZ}_{0+}$ Coefficient |
|---|---|
| $Cos[\omega_\oplus T_\oplus]$ | $Cos[\chi]\left((\mathcal{M}_{DB})^{yz}_{lab} - (\mathcal{M}_{DB})^{zy}_{lab}\right) + Sin[\chi]\left((\mathcal{M}_{DB})^{xy}_{lab} - (\mathcal{M}_{DB})^{yx}_{lab}\right)$ |
| $Sin[\omega_\oplus T_\oplus]$ | $\left((\mathcal{M}_{DB})^{xz}_{lab} - (\mathcal{M}_{DB})^{zx}_{lab}\right)$ |

Here $\chi$ is the angle of the colatitude of the experiment from the north-pole of the earth, $\eta$ is the angle between the celestial equatorial plane and the ecliptic, the path of the Earth's orbital motion (~$23.4^0$), $\Omega_\oplus$ and $v_\oplus$ are the angular frequency and speed of the Earth's orbital motion, $\omega_\oplus$ is the angular frequency of the Earth's rotation, $T$ is the time since a spring equinox, and $T_\oplus$ is the time since the laboratory frame $y$-axis pointed towards 90 ° right ascension.

## 2. Rotating laboratory experiments

To calculate the dependence of a rotating laboratory experiment, we assume rotation is taken about the z-axis of the laboratory frame at a spin frequency of $\omega_s$. As suggested in Kostelecky and Mewes,[11] we transform the $\mathcal{M}$ matrices using a standard tensor rotations in the laboratory frame, so that the components of the $\mathcal{M}$ matrices become time dependent at the spin frequency and twice the spin frequency. Of course the $\mathcal{M}^{ZZ}$ components are the only ones to remain unaffected. Also, it should be noted that through the rotation the $\tilde{\kappa}^{xy}_{0+}$ can be directly measured without the boost term suppression. Here we define $v_{equ}/c$ as the speed of the earth's rotation at the equator expressed as a fraction of the speed of light, which is $1.5\times10^{-6}$, and the time $T_S$ is the time since the rotating experiment's $y$-axis pointed towards 90° right ascension.

**Table IV, Sensitivity coefficients of $(\mathrm{v}_\oplus/c)\tilde{\kappa}_{tr}$ at the relevant modulation frequencies**

| Modulation | $(\mathrm{v}_\oplus/c)\tilde{\kappa}_{tr}$ Coefficient |
|---|---|
| $Cos[\omega_S T_S]$ | $-2\frac{\mathrm{v}_{equ}}{\mathrm{v}_\oplus} Sin[\chi]\left((\mathcal{M}_{DB})^{xz}_{lab} - (\mathcal{M}_{DB})^{zx}_{lab}\right)$ |
| $Sin[\omega_S T_S]$ | $-2\frac{\mathrm{v}_{equ}}{\mathrm{v}_\oplus} Sin[\chi]\left((\mathcal{M}_{DB})^{yz}_{lab} - (\mathcal{M}_{DB})^{zy}_{lab}\right)$ |
| $Cos[\omega_S T_S + \Omega_\oplus T]$ | $Sin[\chi]Sin[\eta]\left((\mathcal{M}_{DB})^{yz}_{lab} - (\mathcal{M}_{DB})^{zy}_{lab}\right)$ |
| $Sin[\omega_S T_S + \Omega_\oplus T]$ | $-Sin[\chi]Sin[\eta]\left((\mathcal{M}_{DB})^{xz}_{lab} - (\mathcal{M}_{DB})^{zx}_{lab}\right)$ |
| $Cos[\omega_S T_S - \Omega_\oplus T]$ | $Sin[\chi]Sin[\eta]\left((\mathcal{M}_{DB})^{yz}_{lab} - (\mathcal{M}_{DB})^{zy}_{lab}\right)$ |
| $Sin[\omega_S T_S - \Omega_\oplus T]$ | $-Sin[\chi]Sin[\eta]\left((\mathcal{M}_{DB})^{xz}_{lab} - (\mathcal{M}_{DB})^{zx}_{lab}\right)$ |
| $Cos[\omega_S T_S + \omega_\oplus T_\oplus + \Omega_\oplus T]$ | $-\frac{1}{2}(1-Cos[\chi])(1-Cos[\eta])\left((\mathcal{M}_{DB})^{xz}_{lab} - (\mathcal{M}_{DB})^{zx}_{lab}\right)$ |
| $Sin[\omega_S T_S + \omega_\oplus T_\oplus + \Omega_\oplus T]$ | $-\frac{1}{2}(1-Cos[\chi])(1-Cos[\eta])\left((\mathcal{M}_{DB})^{yz}_{lab} - (\mathcal{M}_{DB})^{zy}_{lab}\right)$ |
| $Cos[\omega_S T_S + \omega_\oplus T_\oplus - \Omega_\oplus T]$ | $\frac{1}{2}(1-Cos[\chi])(1+Cos[\eta])\left((\mathcal{M}_{DB})^{xz}_{lab} - (\mathcal{M}_{DB})^{zx}_{lab}\right)$ |
| $Sin[\omega_S T_S + \omega_\oplus T_\oplus - \Omega_\oplus T]$ | $\frac{1}{2}(1-Cos[\chi])(1+Cos[\eta])\left((\mathcal{M}_{DB})^{yz}_{lab} - (\mathcal{M}_{DB})^{zy}_{lab}\right)$ |
| $Cos[\omega_S T_S - \omega_\oplus T_\oplus + \Omega_\oplus T]$ | $\frac{1}{2}(1+Cos[\chi])(1+Cos[\eta])\left((\mathcal{M}_{DB})^{xz}_{lab} - (\mathcal{M}_{DB})^{zx}_{lab}\right)$ |
| $Sin[\omega_S T_S - \omega_\oplus T_\oplus + \Omega_\oplus T]$ | $\frac{1}{2}(1+Cos[\chi])(1+Cos[\eta])\left((\mathcal{M}_{DB})^{yz}_{lab} - (\mathcal{M}_{DB})^{zy}_{lab}\right)$ |
| $Cos[\omega_S T_S - \omega_\oplus T_\oplus - \Omega_\oplus T]$ | $-\frac{1}{2}(1+Cos[\chi])(1-Cos[\eta])\left((\mathcal{M}_{DB})^{xz}_{lab} - (\mathcal{M}_{DB})^{zx}_{lab}\right)$ |
| $Sin[\omega_S T_S - \omega_\oplus T_\oplus - \Omega_\oplus T]$ | $-\frac{1}{2}(1+Cos[\chi])(1-Cos[\eta])\left((\mathcal{M}_{DB})^{yz}_{lab} - (\mathcal{M}_{DB})^{zy}_{lab}\right)$ |

**Table V, Sensitivity coefficients of $\tilde{\kappa}^{xy}_{0+}$ at the relevant modulation frequencies**

| Modulation | $\tilde{\kappa}^{xy}_{0+}$ Coefficient |
|---|---|
| $Cos[\omega_S T_S]$ | $-Sin[\chi]\left((\mathcal{M}_{DB})^{yz}_{lab} - (\mathcal{M}_{DB})^{zy}_{lab}\right)$ |
| $Sin[\omega_S T_S]$ | $Sin[\chi]\left((\mathcal{M}_{DB})^{xz}_{lab} - (\mathcal{M}_{DB})^{zx}_{lab}\right)$ |

**Table VI, Sensitivity coefficients of $\tilde{\kappa}^{xz}_{0+}$ at the relevant modulation frequencies**

| Modulation | $\tilde{\kappa}^{xz}_{0+}$ Coefficient |
|---|---|
| $Cos[\omega_S T_S + \omega_\oplus T_\oplus]$ | $(1-Cos[\chi])\left((\mathcal{M}_{DB})^{xz}_{lab} - (\mathcal{M}_{DB})^{zx}_{lab}\right)$ |
| $Sin[\omega_S T_S + \omega_\oplus T_\oplus]$ | $(1-Cos[\chi])\left((\mathcal{M}_{DB})^{yz}_{lab} - (\mathcal{M}_{DB})^{zy}_{lab}\right)$ |
| $Cos[\omega_S T_S - \omega_\oplus T_\oplus]$ | $(1+Cos[\chi])\left((\mathcal{M}_{DB})^{xz}_{lab} - (\mathcal{M}_{DB})^{zx}_{lab}\right)$ |
| $Sin[\omega_S T_S - \omega_\oplus T_\oplus]$ | $(1+Cos[\chi])\left((\mathcal{M}_{DB})^{yz}_{lab} - (\mathcal{M}_{DB})^{zy}_{lab}\right)$ |

**Table VII, Sensitivity coefficients of $\tilde{\kappa}_{0+}^{yz}$ at the relevant modulation frequencies**

| Modulation | $\tilde{\kappa}_{0+}^{yz}$ Coefficient |
|---|---|
| $Cos[\omega_S T_S + \omega_\oplus T_\oplus]$ | $-(1-Cos[\chi])\left((\mathcal{M}_{DB})_{lab}^{yz} - (\mathcal{M}_{DB})_{lab}^{zy}\right)$ |
| $Sin[\omega_S T_S + \omega_\oplus T_\oplus]$ | $(1-Cos[\chi])\left((\mathcal{M}_{DB})_{lab}^{xz} - (\mathcal{M}_{DB})_{lab}^{zx}\right)$ |
| $Cos[\omega_S T_S - \omega_\oplus T_\oplus]$ | $(1+Cos[\chi])\left((\mathcal{M}_{DB})_{lab}^{yz} - (\mathcal{M}_{DB})_{lab}^{zy}\right)$ |
| $Sin[\omega_S T_S - \omega_\oplus T_\oplus]$ | $-(1+Cos[\chi])\left((\mathcal{M}_{DB})_{lab}^{xz} - (\mathcal{M}_{DB})_{lab}^{zx}\right)$ |

## APPENDIX B: RECTANGULAR WAVEGUIDE IN THE SME

In this appendix we consider a perfectly conducting rectangular waveguide, filled with isotropic material of relative permittivity and permeability $\varepsilon_r$ and $\mu_r$ respectively, with height $a$ and width $b$, oriented along the z-axis in the laboratory frame. The solution of Maxwell's equations using separation of variable gives the following well-known solution for Transverse Electric (TE) modes.

$$\boldsymbol{E} = \left(E_{xo}\hat{\boldsymbol{x}} + E_{yo}\hat{\boldsymbol{y}} + E_{zo}\hat{\boldsymbol{z}}\right)e^{-i\beta z}e^{i\omega t} \qquad \text{(B1)}$$
$$\boldsymbol{B} = \left(B_{xo}\hat{\boldsymbol{x}} + B_{yo}\hat{\boldsymbol{y}} + B_{zo}\hat{\boldsymbol{z}}\right)e^{-i\beta z}e^{i\omega t}$$

$$B_{zo} = B_o \cos[k_x x]\cos[k_y y]$$

$$B_{xo} = B_o \frac{i\beta.k_x}{k^2-\beta^2}\sin[k_x x]\cos[k_y y]$$

$$B_{yo} = B_o \frac{i\beta.k_y}{k^2-\beta^2}\cos[k_x x]\sin[k_y y]$$

$$E_{xo} = B_o \frac{i\omega.k_y}{k^2-\beta^2}\cos[k_x x]\sin[k_y y]$$

$$E_{yo} = B_o \frac{-i\omega.k_x}{k^2-\beta^2}\sin[k_x x]\cos[k_y y]$$

where $k=\omega\sqrt{\varepsilon_0\mu_0\varepsilon_r\mu_r}$ , is the free space propagation constant, $k_x = \frac{m\pi}{a}$ and $k_y = \frac{n\pi}{b}$, where $n$ and $m$ are integers greater or equal to zero. To find the phase shift over length $L$ due to leading order Lorentz violating terms in the SME we compute the $\mathcal{M}$ matrices given by (28), and find the following non-zero components;

$$(\mathcal{M}_{DE})_{lab}^{xx} = \frac{\pi L}{\lambda_v}\left(\frac{k_y^2}{k_x^2+k_y^2}\right)\frac{v_{ph}}{c}\mu_r \qquad \text{(B2)}$$
$$(\mathcal{M}_{DE})_{lab}^{yy} = \frac{\pi L}{\lambda_v}\left(\frac{k_x^2}{k_x^2+k_y^2}\right)\frac{v_{ph}}{c}\mu_r$$

$$\left(\mathcal{M}_{HB}\right)_{lab}^{xx} = -\frac{\pi L}{\lambda_v}\left(\frac{k_x^2}{k_x^2 + k_y^2}\right)\frac{c}{v_{ph}}\mu_r$$

$$\left(\mathcal{M}_{HB}\right)_{lab}^{yy} = -\frac{\pi L}{\lambda_v}\left(\frac{k_y^2}{k_x^2 + k_y^2}\right)\frac{c}{v_{ph}}\mu_r$$

$$\left(\mathcal{M}_{HB}\right)_{lab}^{zz} = -\frac{\pi L}{\lambda_v}\left(\frac{k_x^2 + k_y^2}{k_0^2}\right)\frac{v_{ph}}{c}\mu_r$$

$$\left(\mathcal{M}_{DB}\right)_{lab}^{xy} = \frac{2\pi L}{\lambda_v}\left(\frac{k_y^2}{k_x^2 + k_y^2}\right)\mu_r$$

$$\left(\mathcal{M}_{DB}\right)_{lab}^{yx} = -\frac{2\pi L}{\lambda_v}\left(\frac{k_x^2}{k_x^2 + k_y^2}\right)\mu_r$$

$$v_{ph} = \frac{\omega}{\sqrt{k^2 - k_x^2 - k_y^2}}$$

Here, $v_{ph}$ is the phase velocity of the mode inside the waveguide. These values reduce to those calculated for the plane wave in the appropriate limits. By letting $a$ and then $b$ tend to infinity we recover the ($x$, $y$) polarization and vice versa for the ($y$, $x$) polarization . It should be noted that;

$$\left(\mathcal{M}_{DB}\right)_{lab}^{xy} - \left(\mathcal{M}_{DB}\right)_{lab}^{yx} = \frac{2\pi L}{\lambda_v}\mu_r \qquad \text{(B3)}$$

is independent of $k_x$ and $k_y$, and is the same as the plane wave case. Because equation (B3) is non-zero, the propagation of the mode will depend on the $\tilde{\kappa}_{tr}$ and $\tilde{\kappa}_{0+}$coefficients in the sun-centered celestial equatorial frame as shown in Appendix A. Furthermore, the sensitivity is independent of mode of propagation in the waveguide.